\documentclass[onecolumn]{aa}  
\usepackage{graphicx}
\usepackage{txfonts}
\usepackage{url}
\begin{document}

\title{``TNOs are Cool'': A survey of the trans-Neptunian region}

\subtitle{VI. \emph{Herschel}\thanks{{\it Herschel} is an ESA space observatory with science instruments provided by
European-led Principal Investigator consortia and with important participation from NASA.}/PACS observations and thermal
modeling of 19 classical Kuiper belt objects}

\author{E. Vilenius\inst{1} \and C. Kiss\inst{2} \and M. Mommert\inst{3} \and T. M\"uller\inst{1} \and
P. Santos-Sanz\inst{4} \and A. Pal\inst{2} \and
J. Stansberry\inst{5} \and M. Mueller\inst{6,7} \and N. Peixinho\inst{8,9}
S. Fornasier\inst{4,10} \and E. Lellouch\inst{4} 
\and A. Delsanti\inst{11} \and 
A. Thirouin\inst{12} \and J.~L.~Ortiz\inst{12} \and R. Duffard\inst{12} \and
D. Perna\inst{13,14} \and N. Szalai\inst{2} \and 
S. Protopapa\inst{15} \and F. Henry\inst{4} \and D. Hestroffer\inst{16} 
\and M. Rengel\inst{17} \and E. Dotto\inst{13} \and P. Hartogh\inst{17}
}

\institute{Max-Planck-Institut f\"ur extraterrestrische Physik, Postfach 1312, Giessenbachstr., 85741 Garching, Germany\\
\email{vilenius@mpe.mpg.de}
\and
Konkoly Observatory of the Hungarian Academy of Sciences, 1525 Budapest, PO~Box~67, Hungary
\and
Deutsches Zentrum f\"ur Luft- und Raumfahrt e.V., Institute of Planetary Research, Rutherfordstr. 2, 12489 Berlin, Germany
\and
LESIA-Observatoire de Paris, CNRS, UPMC Univ. Paris 06, Univ. Paris-Diderot, France
\and
Stewart Observatory, The University of Arizona, Tucson AZ 85721, USA
\and
SRON LEA / HIFI ICC, Postbus 800, 9700AV Groningen, Netherlands
\and
UNS-CNRS-Observatoire de la C{\^o}te d'Azur, Laboratoire Cassiope\'e, BP 4229, 06304 Nice Cedex 04, France
\and
Center for Geophysics of the University of Coimbra, Av.~Dr.~Dias da Silva, 3000-134 Coimbra, Portugal
\and
Astronomical Observatory of the University of Coimbra, Almas de Freire, 3040-04 Coimbra, Portugal
\and
Univ. Paris Diderot, Sorbonne Paris Cit\'{e}, 4 rue Elsa Morante, 75205 Paris, France
\and
Laboratoire d'Astrophysique de Marseille, CNRS \& Universit\'e de Provence, 38 rue Fr\'ed\'eric Joliot-Curie,
13388 Marseille Cedex 13, France
\and
Instituto de Astrof\'isica de Andaluc\'ia (CSIC), Camino Bajo de Hu\'etor 50, 18008 Granada, Spain
\and
INAF -- Osservatorio Astronomico di Roma, via di Frascati, 33, 00040 Monte Porzio Catone, Italy
\and
INAF - Osservatorio Astronomico di Capodimonte, Salita Moiariello 16, 80131 Napoli, Italy
\and
University of Maryland, College Park, MD 20742, USA
\and
IMCCE, Observatoire de Paris, 77 av.~ Denfert-Rocherea, 75014, Paris, France
\and
Max-Planck-Institut f\"ur Sonnensystemforschung, Max-Planck-Stra{\ss}e 2, 37191 Katlenburg-Lindau, Germany
}

\date{Received December 26, 2011; accepted March 6, 2012}

\abstract
{Trans-Neptunian objects (TNO) represent the leftovers of the formation of the Solar System. Their physical
properties provide constraints to the models of formation and evolution of the various dynamical classes of
objects in the outer Solar System.}
{Based on a sample of 19 classical TNOs we determine radiometric sizes, geometric albedos and
beaming parameters.
Our sample is composed of both dynamically hot and cold classicals. We study the correlations of
diameter and albedo of these two subsamples with each other and with orbital parameters, spectral slopes and colors.
}
{We have done three-band photometric observations with \emph{Herschel}/PACS and we
use a consistent method for data reduction and aperture photometry of this sample to obtain monochromatic
flux densities at 70.0, 100.0 and $160.0\ \mathrm{\mu m}$.
Additionally, we use \emph{Spitzer}/MIPS flux densities at 23.68 and $71.42\ \mathrm{\mu m}$ when available,
and we present new \emph{Spitzer} flux densities of eight 
targets.
We derive diameters and albedos with the near-Earth asteroid thermal model (NEATM). 
As auxiliary data we use reexamined absolute visual magnitudes from the literature and
data bases, part of which have been obtained by ground based
programs in support of our \emph{Herschel} key program.
}
{We have determined for the first time radiometric sizes and albedos of eight classical TNOs,
and refined previous size and albedo estimates or limits of 11 other classicals.
The new size estimates of 2002 MS$_4$ and 120347 Salacia indicate that they are
among the 10 largest TNOs known. 
Our new results confirm the recent findings that there are very diverse albedos
among the classical TNOs and that cold classicals possess a high average albedo ($0.17 \pm 0.04$).
Diameters of classical TNOs strongly correlate with orbital inclination in our sample.
We also
determine the bulk densities of six binary TNOs.}
{}


\maketitle

\section{Introduction}
The physical properties of small Solar System bodies offer constraints
on theories of the formation and evolution of the planets. Trans-Neptunian objects (TNO),
also known as Kuiper Belt objects (KBO), represent
the leftovers from the formation period of the outer Solar System (\cite{Morbidelli2008}),
and they are analogues to the parent bodies
of dust in debris disks around other stars
(\cite[and references therein]{Wyatt2008, MoroMartin2008}).

In addition to Pluto more than 1400 TNOs have been discovered since the first
Kuiper belt object in 1992 (\cite{Jewitt1993}), and the current discovery rate
is 10 to 40 new TNOs/year. 
The dynamical classification is based on the current short-term dynamics.
We use the classification of
Gladman~(\cite{Gladman2008}, $10$~Myr time-scale): classical TNOs are those non-resonant
TNOs which do not belong
to any other TNO class. The classical TNOs are further
divided into the main classical belt,
the inner belt ($a$\,$<$\,$39.4\ \mathrm{AU}$) and the outer belt
($a$\,$>$\,$48.4\ \mathrm{AU}$).
The eccentricity limit for classicals is $e$\,$<$\,$0.24$, beyond which targets are classified
as detached or scattered objects.
The classification scheme of
the Deep Eplictic Survey Team~(DES, \cite{Elliot2005} 2005) differs from the Gladman system in terms of
the boundary of classical and scattered objects, which do not show a clear demarcation
in their orbital parameters. Some of the classicals in the Gladman system are scattered-near
or scattered-extended in the DES system. Another division is 
made in the inclination/eccentricity space. Although there is no dynamical separation, there
seems to be two distinct but partly overlapping inclination distributions 
with the low-i
``cold'' classicals, limited to the main classical belt, showing different average
albedo~(\cite{Grundy2005, Brucker2009} 2009), color~(\cite{Trujillo2002}),
luminosity function~(\cite{Fraser2010} 2010), and frequency
of binary systems~(\cite{Noll2008}) than the high-i ``hot'' classicals,
which has a wider inclination distribution.
Furthermore, models based on recent surveys suggest that there is
considerable sub-structure within the main classical belt (\cite{Petit2011}).
To explain these differences more quantitative data on physical size
and surface composition are needed.

The physical characterization of TNOs has been limited by their large distance and
relatively small sizes. 
Accurate albedos help to correctly interpret
spectra and are needed to find correlations in the albedo-size-color-orbital parameters space
that trace dynamical and collisional history.
The determination of the size frequency distribution (SFD) of TNOs
provides one constraint to formation models and gives the total mass.
The SFD of large bodies is dominated by accretion processes and they
hold information about the angular momentum of the pre-solar nebula whereas
bodies smaller than 50 to $100\,\mathrm{km}$ are the result of collisional evolution (\cite{Petit2008}).
The SFD can be estimated 
via the luminosity function (LF), but this
size distribution also depends on assumptions made about surface properties such as albedo.
Consequently, ambiguities in the size distributions derived from the LFs of various
dynamical classes are one significant reason why there is a wide uncertainty in
the total TNO mass estimate ranging from 0.01\,$M_{\mathrm{Earth}}$ (\cite{Bernstein2004})
to 0.2\,$M_{\mathrm{Earth}}$ (\cite{Chiang1999}).
Among the formation models of our Solar System the ``Nice'' family of models  have
been successful in explaining the orbits of planets and the formation of the Kuiper belt
(\cite{Tsiganis2005}, \cite{Levison2008}), although 
they have difficulties in explaining some of the details of the cold and hot distributions
and the origin of the two sub-populations
(e.g.~\cite{Fraser2010} 2010, \cite{Petit2011,Batygin2011}).

Only a few largest TNOs have optical size estimates based
on direct imaging and assumptions about the limb darkening function
(e.g. Quaoar, \cite{Fraser2010b} 2010).
The combination of optical and thermal infrared observations gives both sizes and geometric albedos,
but requires thermal modeling. Earlier results from \emph{Spitzer} and \emph{Herschel} have shown the usefulness of this
method (e.g. \cite{Stansberry2008} 2008, \cite{Muller2010} 2010) and significantly changed the size and albedo estimates of several TNOs
compared to those obtained by using an assumed albedo.

In this work we present new radiometric diameters and geometric albedos for 19 classical TNOs.
Half of them have no previously published observations in the wavelength regime used in this work.
Those which have been observed before by \emph{Spitzer} now have more complete sampling
of their SEDs close to the thermal peak.
The new estimates of the 19 targets are based on
observations performed with the ESA \emph{Herschel Space Observatory} (\cite{Pilbratt2010}) and its
Photodetector Array Camera and Spectrometer (PACS; \cite{Poglitsch2010}). 
Other \emph{Herschel} results for TNOs have been presented by \cite{Muller2010} (2010), \cite{Lellouch2010}
(2010) and \cite{Lim2010} (2010).
New estimates of 18 Plutinos are presented in \cite{Mommert2012} (2012) and
of 15 scattered disc and detached objects in \cite{SantosSanz2012} (2012).

This paper is organized in the following way. We describe our target sample in Section~\ref{Targetsample},
\emph{Herschel} observations in Section~\ref{Hobs} and \emph{Herschel} data reduction in Section~\ref{dataredux}.
New or re-analyzed flux densities from \emph{Spitzer} are presented in Section~\ref{Sobs}. As auxiliary data we
use absolute V-band magnitudes (Section~\ref{auxobs}), which we have adopted from other works or data
bases taking into account the factors relevant to their uncertainty estimates.
Thermal modeling is described in
Section~\ref{model} and the results for individual targets in Section~\ref{resultsection}.
In Section~\ref{discussions} we discuss sample properties of our sample and of all classicals with radiometric diameters
and geometric albedos as well as correlations (Section~\ref{correlations}) and the bulk densities of binaries (Section~\ref{binaries}).
Finally, the conclusions are in Section~\ref{conclude}.

\section{Observations and data reduction}
Our sample of 19 TNOs 
has been observed as part
of the {\it Herschel} key program
``TNOs are Cool'' (\cite{Muller2009}) mainly between February and November 2010 by the photometry sub-instrument
of PACS in the wavelength range
60--$210\ \mathrm{\mu m}$.

\subsection{Target sample}
\label{Targetsample}
The target sample consists of both dynamically cold and hot classicals (Table~\ref{table_overview}).
We use a cut-off limit of $i=4.5\degr$ in illustrating the two subsamples.
Another typical value used in the literature is $i=5\degr$. The inclination limit is lower
for large objects (\cite{Petit2011}) which have a higher probability of belonging to the
hot population.
Targets 119951 (2002 KX$_{14}$), 120181 (2003 UR$_{292}$) and 78799 (2002 XW$_{93}$)
are in the inner classical belt and are therefore considered to belong to the low-inclination tail
of the hot population. The latter two would be Centaurs in the DES system
and all targets
in Table~\ref{table_overview} with $i$\,$>$\,$15\degr$ would belong to the scattered-extended class of DES.

The median absolute V-magnitudes ($H_{\mathrm{V}}$, see Section~\ref{auxobs}) of our sample are 6.1\, mag
for the cold sub-sample and 5.3\, mag for the hot one.
\cite{Levison2001} (2001) found that bright classicals
have systematically higher inclinations than fainter ones.  
This trend is seen among our targets 
(see Section~\ref{corr_other}).
Another known population characteristics is the lack of a
clear color demarcation line at $i\approx5\degr$ (\cite{Peixinho2008}), 
which is absent also from our sample of 10 targets with known colors.
\begin{table*}
\centering
\caption{Target sample. Semimajor axis $a$, perihelion distance $q$, inclination $i$, eccentricity $e$,
color taxonomy, spectral slope,
and the average absolute visual magnitudes (V or R-band) of 19 TNOs ordered according to
increasing inclination. For $H_\mathrm{V}$ used
in our analysis, see Table~\ref{table_overview_groundobs}.}
\begin{tabular}{lcccccclc}
\hline
\hline
Target & KBO & $a$ \tablefootmark{a} & $q$  & $i$ \tablefootmark{a}    & $e$ \tablefootmark{a} & Color \tablefootmark{b} & Spectral slope & $H_\mathrm{V}$\\
       & location &(AU)                  & (AU) & (\degr)                  &                      &                      & (\% / 100 nm)  & (mag) \\
\hline
\object{119951 (2002 KX$_{14}$)} & {\bfseries inner} & 38.9 & 37.1 & 0.4 & 0.05 & RR-IR\tablefootmark{c} & $27.1 \pm 1.0$\tablefootmark{e} & $4.862 \pm 0.038$\tablefootmark{r} \\
\object{(2001 XR$_{254}$)}$^*$            & main & 43.0 & 41.7 & 1.2 & 0.03 & ... & $10 \pm 3$\tablefootmark{f} & $6.030 \pm 0.017$\tablefootmark{s} \\
\object{275809 (2001 QY$_{297}$)}$^*$    & main & 44.0 & 40.4 & 1.5 & 0.08 & BR & $24 \pm 8$\tablefootmark{f,g,h} & $6.09 \pm 0.03$\tablefootmark{f} \\
\object{(2001 RZ$_{143}$)}$^*$	   & main & 44.4 & 41.3 & 2.1 & 0.07 & ... & $13 \pm 6$\tablefootmark{h,i,j} & $6.69 \pm 0.10$\tablefootmark{i}  \\
\object{(2002 GV$_{31}$)}	         & main  & 43.9 &     40.0 &  2.2 & 0.09 & ... & \ldots & $H_\mathrm{R}=5.5$\tablefootmark{t} $\pm 0.4 $ \\
\object{79360 Sila (1997 CS$_{29}$)}$^*$ & main & 43.9 & 43.4 & 2.2 & 0.01 & RR & $27.0 \pm 3.0$\tablefootmark{e} & $5.59 \pm 0.06$\tablefootmark{f,u} \\
\object{88611 Teharonhiawako (2001 QT$_{297}$)}$^*$ & main & 44.2 & 43.2 &  2.6 & 0.02 & ... & ~~$1 \pm 2$\tablefootmark{f} & $5.97 \pm 0.03$\tablefootmark{f} \\
\object{120181 (2003 UR$_{292}$)} & {\bfseries inner} & 32.6 & 26.8 & 2.7 & 0.18 & ... & $28 \pm 5$\tablefootmark{k} & $H_\mathrm{R}=6.7$\tablefootmark{t} $\pm 0.3$ \\
\object{(2005 EF$_{298}$)}                                 & main  & 43.9 &     40.1 &  2.9  & 0.09 & RR & \ldots & $H_\mathrm{R}=5.8$\tablefootmark{t} $\pm 0.3 $ \\
\hline
\object{138537 (2000 OK$_{67}$)}            & main & 46.8 & 40.0 & 4.9 & 0.14 & RR & $20 \pm 3$\tablefootmark{h,j,l,m,n} & $6.47 \pm 0.09$\tablefootmark{l,m}  \\
\object{148780 Altjira (2001 UQ$_{18}$)}$^*$ & main & 44.5 & 41.8 &  5.2 & 0.06 & RR & $35 \pm 6$\tablefootmark{f,g,h,j} & $6.47 \pm 0.13$\tablefootmark{f,g}  \\
\object{(2002 KW$_{14}$)}	                           & main & 46.5 &      37.3 &  9.8  & 0.20 & ...   & \ldots & $5.88 \pm 0.05$\tablefootmark{v} \\
\object{(2001 KA$_{77}$)}       & main & 47.3 & 42.8 & 11.9 & 0.10 & RR & $38 \pm 3$\tablefootmark{g,h,j,m,o} & $5.64 \pm 0.08$\tablefootmark{g}   \\
\object{19521 Chaos (1998 WH$_{24}$)}  & main & 46.0 & 41.1 & 12.0 & 0.11 & IR & $23 \pm 2$\tablefootmark{h,j,m,n,p,q} & $4.97 \pm 0.05$\tablefootmark{g,u}  \\
\object{78799 (2002 XW$_{93}$)}                   & {\bfseries inner} & 37.6 & 28.3 & 14.3 & 0.25 & ... & \ldots & $H_\mathrm{R}=4.8$\tablefootmark{t} $\pm 0.6$  \\
\object{(2002 MS$_4$)}	               & main & 41.7 & 35.6 & 17.7 & 0.15 & ... & ~~$2 \pm 2$\tablefootmark{k} & $H_\mathrm{R}=3.5$\tablefootmark{t} $\pm 0.4$ \\
\object{145452 (2005 RN$_{43}$)} & main & 41.8 & 40.6 & 19.2 & 0.03 & RR-IR\tablefootmark{c} & $23.0 \pm 1.1$\tablefootmark{e} & $3.89 \pm 0.05$\tablefootmark{v} \\
\object{90568 (2004 GV$_9$)}     & main & 41.8 & 38.7 & 22.0 & 0.07 & BR\tablefootmark{d} & $15 \pm 3$\tablefootmark{k} & $4.25 \pm 0.04$\tablefootmark{w}      \\
\object{120347 Salacia (2004 SB$_{60}$)}$^*$ & main & 42.2 & 37.9 & 23.9 & 0.10 & ... & $12.6 \pm 2.0$\tablefootmark{e} & $4.26 \pm 0.02$\tablefootmark{f}  \\
\hline
\end{tabular}
\label{table_overview}
\tablefoot{The horizontal line marks the limit of dynamically cold and
hot classicals according to our dynamical analysis using the Gladman system.
* denotes a known binary system (\cite{Noll2008}). {\bf References}.
\tablefoottext{a}{IAU Minor Planet Center, \url{http://www.minorplanetcenter.net/iau/Ephemerides/Distant/}, accessed July 2011.}
\tablefoottext{b}{Taxonomic class from \cite{Fulchignoni2008} unless otherwise indicated.}
\tablefoottext{c}{\cite{Barucci2011} (2011).}
\tablefoottext{d}{\cite{Perna2010}.}
\tablefoottext{e}{\cite{Fornasier2009}.}
\tablefoottext{f}{\cite{Benecchi2009}.}
\tablefoottext{g}{\cite{Doressoundiram2005}.}
\tablefoottext{h}{Calculated using the technique of \cite{Hainaut2002} (2002).}
\tablefoottext{i}{\cite{SantosSanz2009} (2009).}
\tablefoottext{j}{Values from the Minor Bodies in the Outer Solar System database, 
\url{http://www.eso.org/~ohainaut/MBOSS}.}
\tablefoottext{k}{Stephen Tegler, priv. comm.}
\tablefoottext{l}{\cite{Benecchi2011} (2011).}
\tablefoottext{m}{\cite{Doressoundiram2002}.}
\tablefoottext{n}{\cite{Delsanti2001}.}
\tablefoottext{o}{\cite{Peixinho2004}.}
\tablefoottext{p}{\cite{Barucci2000}.}
\tablefoottext{q}{\cite{Tegler2000}.}
\tablefoottext{r}{\cite{Rabinowitz2007}.}
\tablefoottext{s}{\cite{Grundy2009}.}
\tablefoottext{t}{IAU Minor Planet Center / List of Transneptunian Objects at
\url{http://www.minorplanetcenter.net/iau/lists/TNOs.html}, accessed June 2011.
See Section~\ref{auxobs} for conversion to $H_\mathrm{V}$.}
\tablefoottext{u}{\cite{Romanishin2005} (2005).}
\tablefoottext{v}{Perna et al., {\it in prep.}}
\tablefoottext{w}{\cite{DeMeo2009}.}}
\end{table*}

\subsection{\emph{Herschel} observations}
\label{Hobs}
PACS is an imaging dual band photometer with a rectangular field of view of
$1.75\arcmin$\,$\times$\,$3.5\arcmin$ with full sampling of the
$3.5\ \mathrm{m}$-telescope's point spread function (PSF).
The two detectors are bolometer arrays, the short-wavelength one has
64\,$\times$\,32 pixels and
the long-wavelength one 32\,$\times$\,16 pixels.
In addition, the short-wavelength array has a filter wheel to select between two bands:
60\,--\,$85\ \mathrm{\mu m}$ or 85\,--\,$125\ \mathrm{\mu m}$, whereas the long-wavelength
band is 125\,--\,$210\ \mathrm{\mu m}$. In the PACS photometric system these bands have been
assigned the reference wavelengths $70.0\ \mathrm{\mu m}$, $100.0\ \mathrm{\mu m}$ and
$160.0\ \mathrm{\mu m}$ and they have the names ``blue'', ``green'' and ``red''.
Both bolometers are read-out at $40\ \mathrm{Hz}$ continuously and binned by a factor of four
on-board.

We specified the PACS observation requests (AOR) using the scan-map Astronomical Observation
Template (AOT) in HSpot, a tool provided by the Herschel Science Ground Segment Consortium.
The scan-map mode was selected due to its better overall performance compared to the
point-source mode (\cite{Muller2010} 2010). In this mode the pointing of the telescope is slewed
at a constant speed over parallel lines, or ``legs''. We used 10 scan legs in each AOR,
separated by $4\arcsec$. The length of each leg was $3.0\arcmin$, except for Altjira where it was
$2.5\arcmin$, and the slewing speed was $20\arcsec s^{-1}$. Each one of
these maps was
repeated from two to five times. 

To choose the number of repetitions, i.e. the duration of observations, for our targets we
used the Standard Thermal Model (see Section~\ref{model}) to predict their flux densities
in the PACS bands. Based on earlier \emph{Spitzer} work (\cite{Stansberry2008}, 2008) we adopted
a geometric albedo of 0.08 and a beaming parameter of 1.25 for observation planning purposes.
The predicted thermal fluxes depend on the sizes, which are connected to the assumed geometric
albedo and the absolute V-magnitudes via Equation~(\ref{pA}).
In some
cases the absolute magnitudes used for planning purposes are quite
different (by up to 0.8 mag) from those used for modeling our data as
more recent and accurate visible photometry was taken into account (see Section~\ref{auxobs}).

The PACS scan-map AOR allows the selection of
either the blue or green channel; the red channel data are taken
simultaneously whichever of those is chosen. The sensitivity of the blue
channel is usually limited by instrumental noise, while the red channel
is confusion-noise limited (\cite{PACSrelnote}). The sensitivity in the green channel can
be dominated by either source, depending on the depth and the region of the sky
of the observation. For a given channel selection (blue or green)
we grouped pairs of AORs, with scan orientations of $70\degr$ and $110\degr$ with respect
to the detector array, in order to
make optimal use of the rectangular shape of the detector. Thus, during a single visit of a
target we grouped 4 AORs to be observed in sequence:
two AORs in different scan directions and this repeated for the second channel selection.

The timing of the observations, i.e.~the selection of the
visibility window, has been optimized to utilize the lowest far-infrared confusion noise
circumstances~(\cite{Kiss2005}) such that the estimated signal-to-noise ratio due to
confusion noise has its maximum in the green channel. Each target was visited twice with
similar AORs repeated in both visits for the purpose of background subtraction.
The timing of the second visit was calculated such
that the target has moved 30-$50\arcsec$ between the visits so that the target position
during the second visit is within the high-coverage area of the map from the first visit.
Thus, we can determine the background for the two source positions.

The observational details are listed in Table~\ref{table_obs}. 
All of the targets observed had predicted astrometric $3\,\sigma$ uncertainties 
less than $10\arcsec$ at the time of the \emph{Herschel} observations (David Trilling, {\it priv.~comm.}).

\begin{table*}
\centering
\caption{Individual observations of the sample of 19 TNOs by {\it Herschel}/PACS.
OBSIDs are the observation identifiers, duration is the total duration of the
four AORs (see text), mid-time is the mean UT time, $r$ is the mean heliocentric
distance, $\Delta$ is the mean {\it Herschel}-target distance, and $\alpha$ is
the mean phase angle. Each line corresponds to one visit and all three channels
were observed at each visit.}
\begin{tabular}{llcrccc}
\hline
Target & OBSIDs  & Duration & Mid-time & $r$  & $\Delta$ & $\alpha$ \\
       &         & (min)    &          & (AU) & (AU)     & (\degr)  \\
\hline
119951 (2002 KX$_{14}$)        &  1342205144-5147 & 59.8 & 26-Sep-2010 21:52:54 & 39.3993 & 39.8605 & 1.30 \\
119951 (2002 KX$_{14}$)        &  1342205175-5178 & 59.8 & 27-Sep-2010 15:35:54 & 39.3993 & 39.8716 & 1.29 \\
(2001 XR$_{254}$)       &  1342205184-5187 & 78.6 & 27-Sep-2010 18:48:10 & 44.1004 & 44.4032 & 1.25 \\
(2001 XR$_{254}$)       &  1342205264-5267 & 78.6 & 28-Sep-2010 15:48:21 & 44.1004 & 44.3886 & 1.25 \\
(2001 QY$_{297}$)       &  1342209492-9495 & 97.4 & 18-Nov-2010 10:06:17 & 43.2452 & 43.3599 & 1.31 \\
(2001 QY$_{297}$)       &  1342209650-9653 & 97.4 & 19-Nov-2010 20:50:15 & 43.2455 & 43.3850 & 1.31 \\
(2001 RZ$_{143}$)	 &  1342199503-9506 & 97.4 & 01-Jul-2010 00:52:31 & 41.3006 & 41.6947 & 1.31 \\
(2001 RZ$_{143}$)	 &  1342199614-9617 & 97.4 & 01-Jul-2010 20:28:08 & 41.3005 & 41.6819 & 1.32 \\
(2002 GV$_{31}$)	 &  1342198847-8850 & 59.8 & 20-Jun-2010 20:07:56 & 40.2818 & 40.5238 & 1.42 \\
(2002 GV$_{31}$)	 &  1342198897-8900 & 59.8 & 21-Jun-2010 21:55:30 & 40.2817 & 40.5412 & 1.41 \\
79360 Sila\tablefootmark{a} & 1342187073 & 94.4 & 18-Nov-2009 14:24:02  & 43.5090 & 43.2410 & 1.27 \\
79360 Sila	 &  1342196137-6140 & 59.8 & 09-May-2010 01:29:23 & 43.5057 & 43.6530 & 1.33 \\
79360 Sila	 &  1342196137-6140 & 59.8 & 10-May-2010 09:15:12 & 43.5057 & 43.6753 & 1.32 \\
88611 Teharonhiawako	 &  1342196099-6102 & 78.6 & 09-May-2010 22:19:55 & 45.0856 & 45.3612 & 1.24 \\
88611 Teharonhiawako	 &  1342196145-6148 & 78.6 & 10-May-2010 18:30:58 & 45.0857 & 45.3475 & 1.25 \\
120181 (2003 UR$_{292}$) &  1342199618-9621 & 59.8 & 01-Jul-2010 21:49:00 & 26.7872  &	27.1477  &	2.04 \\
120181 (2003 UR$_{292}$) &  1342199646-9649 & 59.8 & 02-Jul-2010 12:43:03 & 26.7872  &	27.1379  &	2.05 \\
(2005 EF$_{298}$)       &  1342208962-8965 & 97.4 & 03-Nov-2010 20:07:09 & 40.7207 & 41.1758 & 1.25 \\
(2005 EF$_{298}$)       &  1342208999-9002 & 97.4 & 04-Nov-2010 10:08:12 & 40.7207 & 41.1667 & 1.25 \\
\hline
138537 (2000 OK$_{67}$)	 &  1342197665-7668 & 78.6 & 03-Jun-2010 00:55:51 & 40.3008 & 40.3581 & 1.46 \\
138537 (2000 OK$_{67}$)	 &  1342197717-7720 & 78.6 & 04-Jun-2010 13:55:32 & 40.3006 & 40.3319 & 1.46 \\
148780 Altjira		 &  1342190917-0920 & 76.0 & 21-Feb-2010 23:32:17 & 45.5387 & 45.5571 & 1.25 \\
148780 Altjira		 &  1342191120-1123 & 76.0 & 24-Feb-2010 01:31:06 & 45.5390 & 45.5940 & 1.25 \\
(2002 KW$_{14}$)	 &  1342204196-4199 & 59.8 & 09-Sep-2010 11:11:36 & 40.8385 & 41.0710 & 1.38 \\
(2002 KW$_{14}$)	 &  1342204282-4285 & 59.8 & 10-Sep-2010 09:37:01 & 40.8389 & 41.0870 & 1.38 \\
(2001 KA$_{77}$)	 &  1342205962-5965 & 78.6 & 06-Oct-2010 21:09:43 & 48.1789 & 48.6245 & 1.07 \\
(2001 KA$_{77}$)	 &  1342206013-6016 & 78.6 & 07-Oct-2010 14:32:45 & 48.1787 & 48.6354 & 1.07 \\
19521 Chaos		&  1342202285-2288 & 41.0 & 08-Aug-2010 19:40:03 & 41.6914 & 42.1566 & 1.24 \\
19521 Chaos		&  1342202316-2319 & 41.0 & 09-Aug-2010 11:34:49 & 41.6913 & 42.1464 & 1.25 \\
78799 (2002 XW$_{93}$) & 1342190913-0916  &	57.8  &	21-Feb-2010 22:20:43  &	44.6271  &	44.3163  &	1.22 \\
78799 (2002 XW$_{93}$) & 1342191116-1119  &	57.8  &	24-Feb-2010 00:19:32  &	44.6279  &	44.3518  &	1.23 \\
(2002 MS$_4$)	 &  1342204140-4143 & 41.0 & 08-Sep-2010 21:12:31 & 47.1673 & 46.9025 & 1.19 \\
(2002 MS$_4$)	 &  1342204292-4295 & 41.0 & 10-Sep-2010 15:03:37 & 47.1670 & 46.9307 & 1.20 \\
145452 (2005 RN$_{43}$)        &  1342195583-5586 & 41.0 & 25-Apr-2010 22:21:44 & 40.6885 & 41.1522 & 1.26 \\
145452 (2005 RN$_{43}$)        &  1342195600-5603 & 41.0 & 26-Apr-2010 21:43:12 & 40.6885 & 41.1378 & 1.28 \\
90568 (2004 GV$_9$)         &  1342202869-2872 & 41.0 & 11-Aug-2010 18:35:31 & 39.1876 & 39.3841 & 1.46 \\
90568 (2004 GV$_9$)         &  1342202921-2924 & 41.0 & 12-Aug-2010 15:22:58 & 39.1877 & 39.3984 & 1.46 \\
120347 Salacia          &  1342199133-9136 & 41.0 & 22-Jun-2010 01:17:20 & 44.1464 & 44.0169 & 1.33 \\
120347 Salacia          &  1342199133-9136 & 41.0 & 22-Jun-2010 18:57:52 & 44.1466 & 44.0058 & 1.32 \\
\hline
\end{tabular}
\label{table_obs}
\tablefoot{$r$, $\Delta$ and $\alpha$ are from the JPL Horizons Ephemeris System (\cite{Giorgini1996}).
\tablefoottext{a}{Observation in the chopped and nodded point-source mode and only at the blue and red
channels (\cite{Muller2010} 2010).}}
\end{table*}

\subsection{Data reduction}
\label{dataredux}
\begin{figure*}
   \centering
   \includegraphics[width=17cm]{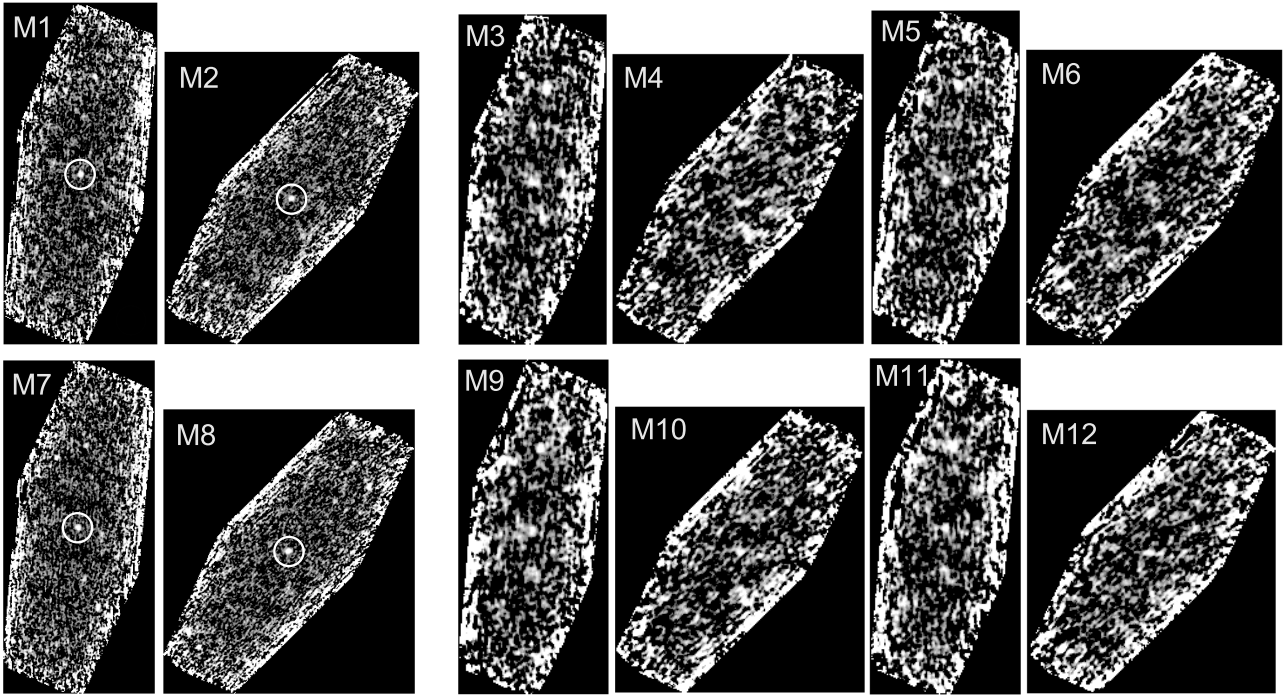}
   \caption{Individual maps of 120347 Salacia. Each map is the product of one observation (AOR).
   The first row (M1-M6) is from the first visit and the second row (M7-M12) from the follow-on visit.
   The first two columns (M1-M2, M7-M8) are observations in the 
   $100\ \mathrm{\mu m}$ or ``green'' channel and the others in the
   $160\ \mathrm{\mu m}$ or ``red'' channel. The two scan angles are 110\degr (odd-numbered maps) and 70\degr.
   The source is clearly seen in the map center in the green
   channel whereas the red channel is more affected by background sources and confusion noise.
   Orientation: north is up and east is to the left.}
   \label{individual_maps}
\end{figure*}

\begin{figure}
\centering
\includegraphics[width=15cm]{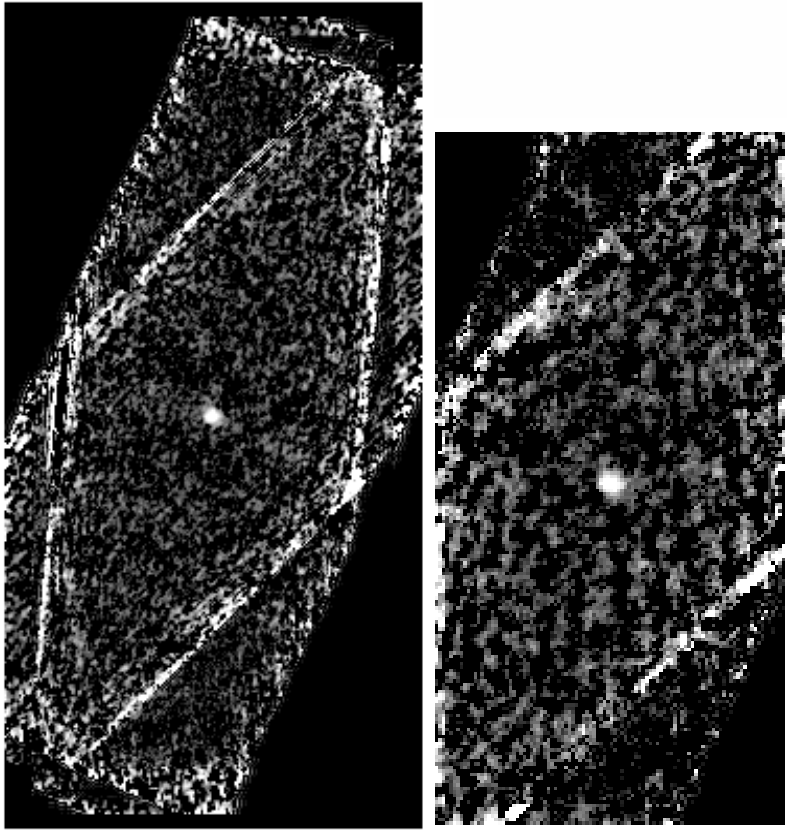}
\caption{Combined maps of 120347 Salacia from the individual maps 
(Fig.~\ref{individual_maps}) in the green (left) and red
channels. Orientation: north is up and east is to the left.}
\label{combined_maps}
\end{figure}

The data
reduction from level 0 (raw data) to level 2 (maps) was done using
Herschel Interactive Processing Environment (HIPE\footnote{Data presented in this paper were
analysed using ``HIPE'', a joint development
by the Herschel Science Ground Segment Consortium, consisting of ESA, the 
NASA Herschel Science Center, and the HIFI, PACS and SPIRE consortia 
members, see \url{http://herschel.esac.esa.int/DpHipeContributors.shtml}.}) with modified scan-map pipeline
scripts optimized for the 
 ``TNOs are Cool'' key program. 
The individual maps (see Fig.~\ref{individual_maps} for examples) from the same epoch and channel are
mosaicked, and background-matching and source-stacking techniques are applied. The two visits
are combined (Fig.~\ref{combined_maps}), in each of the three
bands, and the target with a known apparent motion is located at the center region of these maps.
The detector pixel sizes are $3.2\arcsec \times 3.2\arcsec$
in the blue and green channels, and $6.4\arcsec \times 6.4\arcsec$ in the red channel whereas the pixel sizes
in the maps produced by this data reduction are
$1.1\arcsec$ / $1.4\arcsec$ / $2.1\arcsec$ in the blue / green / red maps, respectively.
A detailed description of the data reduction in the key program is given in Kiss et al. ({\it in prep.}).

Once the target is identified we measure the flux densities at the photocenter position 
using DAOPHOT routines (\cite{Stetson1987}) for aperture photometry.
We make a correction for the encircled energy fraction
of a point source (\cite{PSF2010}) for each aperture used. 
We try to choose the optimum aperture radius in the plateau
of stability of the growth-curves, which is typically 1.0-1.25 times the full-width-half-maximum of the
PSF ($5.2\arcsec$ / $7.7\arcsec$ / $12.0\arcsec$ in the blue / green / red bands,
respectively). The median aperture radius for targets in the ``TNOs are Cool'' program is 5 pixels in the
final maps (pixel sizes 1.1"/1.4"/2.1" in the blue / green / red maps).
For the uncertainty estimation of the flux density we implant 200 artificial sources in the
map in a region close to the source ($<50\arcsec$) excluding the target itself.
A detailed description of how aperture photometry is implemented in our program
is given in \cite{SantosSanz2012} (2012).

In order to obtain monochromatic flux
density values of targets having a spectral energy distribution different from the
default one color corrections are needed.
In the photometric system of the PACS instrument \emph{flux density} is defined to be
the flux density that a source with
a flat spectrum ($\lambda F_{\lambda}=$constant, where $\lambda$ is the wavelength and
$F_{\lambda}$ is the monochromatic flux) would have at the PACS reference wavelengths (\cite{Poglitsch2010}). 
Instead of the flat default spectrum 
we use a cool black body distribution to calculate
correction coefficients for each PACS band. The filter transmission and bolometer response curves needed
for this calculation are available from HIPE, and we take as black body temperature
the disk averaged day-side temperature calculated iteratively for each target
($\frac{8}{9} \times T_{\mathrm{SS}}$, using STM assumptions from Section~\ref{model},
the Lambertian emission model and the sub-solar temperature from Eq.~\ref{Tss}.)
This calculation yields on the average 0.982 / 0.986 / 1.011 
(flux densities are divided by color correction factors) for the
blue / green / red channels, respectively, with small variation among the
targets of our sample.

The absolute flux density calibration of PACS is based on standard stars and large main belt asteroids
and has the uncertainties of 3\% / 3\% / 5\% for the blue / green / red bands (\cite{PACScal2011}).
We have 
taken
these uncertainties into account in the PACS flux densities used in the modeling, although their
contribution to the total uncertainty is small compared to the signal-to-noise ratio of our
observations. 

The color corrected flux densities are given in Table~\ref{table_Pfluxes}. They were determined from
the combined maps of two visits, in total 4 AORs for the blue and green channels and 8 AORs for the red.
The only exceptions are 19521 Chaos and 90568 (2004 GV$_9$), whose one map 
was excluded from our analysis due to a problem in obtaining reliable photometry from those observations.
The uncertainties in Table~\ref{table_Pfluxes} include the
photometric $1\,\sigma$ and absolute calibration $1\,\sigma$ uncertainties. 17 targets were
detected in at least one PACS channel. The upper limits
are the $1\,\sigma$ noise levels of the maps, including both the instrumental noise and
residuals from the eliminated infrared background confusion noise.
79360 Sila has a flux density which is lower by a factor of three
in the red channel than the one published by \cite{Muller2010} (2010).
We have re-analyzed this earlier chopped/nodded observation
using the latest knowledge on calibration and data reduction and found no significant
change in the flux density values.
As speculated in \cite{Muller2010} (2010) the 2009
single-visit \emph{Herschel}
observation was most probably contaminated by a background source.

\begin{table}
\centering
\caption{Color corrected \emph{Herschel} flux densities of the sample of 19 classical
TNOs from coadded images of two visits. F$_{70}$, F$_{100}$ and F$_{160}$ are the
monochromatic flux densities of the PACS blue / green / red channels.}
\begin{tabular}{lrrr}
\hline
Target &   $F_{70}$  & $F_{100}$  & $F_{160}$      \\
       &   (mJy) &  (mJy) &  (mJy)     \\
\hline
\hline
119951 (2002 KX$_{14}$)       &   $7.4 \pm 0.7$  & $10.2 \pm 1.3$  & $7.2 \pm 1.7$     \\
(2001 XR$_{254}$)      &    $2.5 \pm 0.7$    & $<1.0$           & $<1.4$       \\
275809 (2001 QY$_{297}$) &   $1.1 \pm 1.1$    & $4.2 \pm 0.8$    & $2.4 \pm 1.1$       \\
(2001 RZ$_{143}$)	&   $2.2 \pm 0.7$    & $<1.1$           & $<1.1$       \\
(2002 GV$_{31}$)	&   $<0.8$           & $<1.1$           & $<1.7$              \\
79360 Sila	&   $3.9 \pm 0.8$ & $5.5 \pm 1.1$    & $<5.7$       \\
88611 Teharonhiawako	&   $1.9 \pm 0.7$    & $2.1 \pm 0.9$    & $<2.1$               \\
120181 (2003 UR$_{292}$)      &  $<4.6$           &  $3.9 \pm 1.1$                & $<3.8$ \\
(2005 EF$_{298}$)      &   $1.4 \pm 0.6$    & $1.7 \pm 0.8$    & $<1.9$              \\
\hline
138537 (2000 OK$_{67}$)	&   $<0.9$           & $<0.8$           & $<3.1$       \\
148780 Altjira		&   $3.4 \pm 1.0$    & $4.3 \pm 1.6$    & $<2.1$       \\
(2002 KW$_{14}$)	&   $1.9 \pm 0.8$    & $<3.0$           & $<1.4$              \\
(2001 KA$_{77}$)	&   $<2.1$           & $2.7 \pm 1.0$    & $<1.7$       \\
19521 Chaos &   $9.2 \pm 2.2$\tablefootmark{a} & $11.3 \pm 1.2$  & $10.3 \pm 1.8$      \\
78799 (2002 XW$_{93}$)  &  $17.0 \pm 0.9$           &  $17.8 \pm 1.5$                & $12.9 \pm 1.9$ \\
(2002 MS$_4$)	        &   $26.3 \pm 1.3$ & $35.8 \pm 1.5$ & $21.6 \pm 4.2$    \\
145452 (2005 RN$_{43}$) &   $24.8 \pm 1.3$   & $23.9 \pm 1.8$   & $13.9 \pm 1.9$       \\
90568 (2004 GV$_9$) &   $16.9 \pm 1.0$   & $19.3 \pm 1.9$\tablefootmark{b}   & $18.3 \pm 2.9$      \\
120347 Salacia          &   $30.0 \pm 1.2$ & $37.8 \pm 2.0$ & $28.1 \pm 2.7$   \\
\hline
\end{tabular}
\label{table_Pfluxes}
\tablefoot{\tablefoottext{a}{Observation 1342202317 excluded.}
\tablefoottext{b}{Observation 1342202923 excluded.}}
\end{table}

\subsection{Complementary \emph{Spitzer} observations}
\label{Sobs}
About 75 TNOs and Centaurs in the ``TNOs are Cool'' program were also observed
by the
\emph{Spitzer Space Telescope} (\cite{Werner2004}) using the
Multiband
Imaging Photometer for Spitzer (MIPS; \cite{Rieke2004}). 43
targets were detected at a useful signal-to-noise ratio in both the $24\ \mathrm{\mu m}$
and $70\ \mathrm{\mu m}$ bands of that instrument.
As was done for our \emph{Herschel}
program, many of the Spitzer observations utilized multiple AORs for a
single target, with the visits timed to allow subtraction of background
confusion. The MIPS $24\ \mathrm{\mu m}$ band, when combined with
70-160 $\ \mathrm{\mu m}$ data, can provide very strong constraints on the
temperature of the warmest regions of a TNO.

The absolute calibration, photometric methods and color
corrections for the MIPS data are described in \cite{Gordon2007},
\cite{Engelbracht2007} and \cite{Stansberry2007}. Nominal
calibration uncertaintes are 2\% and 4\% in the $24\ \mathrm{\mu m}$ and $70\ \mathrm{\mu m}$
bands respectively. To allow for additional uncertainties that
may be caused by the sky-subtraction process, application of color
corrections, and the faintness of TNOs relative to the MIPS stellar
calibrators, we adopt uncertainties of 3\% and 6\% as has been done
previously for MIPS TNO data (e.g. \cite{Stansberry2008} 2008, \cite{Brucker2009} 2009).
The
effective monochromatic wavelengths of the two MIPS bands we use are $23.68\ \mathrm{\mu m}$
and $71.42\ \mathrm{\mu m}$. 
With an aperture of
$0.85\ \mathrm{m}$ the telescope-limited spatial resolution is $6\arcsec$ and $18\arcsec$ in the
two bands.

\emph{Spitzer} flux densities of 13 targets overlapping our classical TNO sample
are given in Table~\ref{table_Spitzerobs}.
The new and re-analyzed flux densities are based on re-reduction of the data using updated ephemeris positions.
They sometimes differ by $10\arcsec$ or more
from those used to point \emph{Spitzer}. The ephemeris information is used in
the reduction of the raw $70\ \mathrm{\mu m}$ data, for generating the sky background
images, and for accurate placement of photometric apertures. This is especially
important for the Classical TNOs which are among the faintest objects
observed by \emph{Spitzer}. Results for four targets are previously
unpublished: 275809 (2001 QY$_{297}$), 79360 Sila, 88611 Teharonhiawako, and 19521 Chaos.
As a result of the reprocessing of the data, the fluxes for
119951 (2002 KX$_{14}$), 148780 Altjira,
2001 KA$_{77}$ 
and 2002 MS$_4$ differ from those published in \cite{Brucker2009} (2009) and \cite{Stansberry2008} (2008).

The $70\ \mu m$ bands of PACS and MIPS are overlapping and the flux density values
agree typically within $\pm 5\%$ for the six targets observed by both instruments with SNR $\gtrsim2$.

\begin{table*}
 \centering
 \caption{Complementary \emph{Spitzer} observations. The duration includes the
total time of several visits. Observing epochs lasted 1-14 days.
The other quantities are as in Table~\ref{table_obs}. Targets below the horizontal line have
$i>4.5\degr$.}
 \begin{tabular}{lrlccc|cc|cc}
  \hline
  Target  & Duration & Starting day & $r$ & $\Delta$ & $\alpha$ & \multicolumn{2}{c}{Previous works} & \multicolumn{2}{c}{This work} \\
          & (min)    & of observing       &     &          &          & MIPS F$_{24}$       & MIPS F$_{70}$    & MIPS F$_{24}$ & MIPS F$_{70}$     \\
          &          & epoch              &     &          &          & (mJy)               & (mJy)            & (mJy)         & (mJy)             \\
  \hline
119951    & 128.10  & 2005-08-26  & 39.61 & 39.59 & 1.5 & $0.0786 \pm 0.0082$\tablefootmark{a} & $2.22 \pm 1.45$\tablefootmark{a} & $0.080 \pm 0.013$ & $7.7 \pm 1.7$ \\
119951    & 37.27   & 2006-03-30  & 39.58 & 39.20 & 1.4 & $<0.0363$\tablefootmark{b} & $<3.90$\tablefootmark{b} & $0.083 \pm 0.029$ & \ldots \\
275809  & 594.87  & 2008-11-21  & 42.76 & 42.39 & 1.3 & \ldots & \ldots & $<0.01$             & $<1.5$        \\
(2001 RZ$_{143}$)         & 203.30  & 2004-12-26  & 41.38 & 40.98 & 1.3 & $0.046 \pm 0.008$\tablefootmark{a}   & $<0.7$\tablefootmark{a}         & \ldots & \ldots \\
79360   & 494.70 & 2008-05-18 & 43.52 & 43.30 & 1.3  & \ldots & \ldots & $0.05 \pm 0.01$   & $<2.7$         \\
88611       & 398.42  & 2004-11-04  & 45.00 & 44.67 & 1.2  & \ldots & \ldots & $0.029 \pm 0.010$   & $0.7 \pm 0.6$ \\
\hline
138537    & 257.45 & 2004-11-04   & 40.57 & 40.11 & 1.3 & $0.031 \pm 0.007$\tablefootmark{a}   & $<0.8$\tablefootmark{a}         & \ldots & \ldots \\
148780    & 400.92 & 2006-02-16   & 45.33 & 45.08 & 1.2 & $0.0167 \pm 0.0025$\tablefootmark{a}   & $<0.85$\tablefootmark{a} & $0.020 \pm 0.008$ & $<1.67$ \\
(2002 KW$_{14}$)          & 213.05 & 2005-08-26   & 40.04 & 40.03 & 1.5 & $<0.006$\tablefootmark{a}            & $3.3 \pm 1.1$\tablefootmark{a}  & \ldots & \ldots \\
(2001 KA$_{77}$)          & 400.90 & 2006-03-31   & 48.56 & 48.35 & 1.2 & $0.0077 \pm 0.0023$\tablefootmark{a} & $4.12 \pm 0.81$\tablefootmark{a} & $<0.025$            & $<1.4$        \\
19521                &  66.06 & 2004-09-24   & 42.05 & 41.68 & 1.3  & \ldots & \ldots & \ldots                               & $9.0 \pm 3.0$ \\
(2002 MS$_4$)             & 82.04  & 2006-03-31   & 47.39 & 47.48 & 1.2 & $0.391 \pm 0.022$\tablefootmark{b} & $20.0 \pm 4.1$\tablefootmark{b} & $0.40 \pm 0.02$     & $24.7 \pm 2.9$ \\
90568        & 57.00  & 2005-01-28   & 38.99 & 39.01 & 1.5 & $0.166 \pm 0.010$\tablefootmark{b}   & $17.5 \pm 2.2$\tablefootmark{b} & \ldots & \ldots \\
120347             & 227.62 & 2006-12-03   & 43.82 & 43.39 & 1.2 & $0.546 \pm 0.021$\tablefootmark{c}   & $36.6 \pm 3.7$\tablefootmark{c} & \ldots & \ldots \\
  \hline
 \end{tabular}
\label{table_Spitzerobs}
\tablefoot{The \emph{Spitzer} program IDs are:
55 (for the observations of Chaos), 3229 (Teharonhiawako), 3283 (2002 KX14 in 2006, 2002 MS4, 2004 GV9),
3542 (2002 KX14 in 2005, 2001 RZ143, 2000 OK67, Altjira, 2002 KW14, 2001 KA77),
30081 (Salacia), 40016 (Sila), and 50024 (2001 QY297). {\bf References.}
\tablefoottext{a}{Color corrected MIPS flux density with $1\,\sigma$ measurement
uncertainty from \cite{Brucker2009} (2009) together with quadratically added calibration uncertainties.}
\tablefoottext{b}{\cite{Stansberry2008} (2008), uncertainty calculated 
from SNR and includes quadratically added calibration uncertainties, upper limits shown here are $1\,\sigma$.}
\tablefoottext{c}{\cite{Stansberry2012}.}
}
\end{table*}

\subsection{Optical photometry}
\label{auxobs}
The averages of V-band or R-band absolute magnitude values available in the literature are given in
Table~\ref{table_overview}. The most reliable way of determining them
is to observe a target at multiple phase angles
and over a time span enough to determine lightcurve properties, but such complete
data are available only for 119951 (2002 KX$_{14}$) in our sample.
For all other targets assumptions about the phase behavior have been
made due to the lack of coverage in the range of phase angles.

The IAU (H,G) magnitude system
for photometric phase curve corrections (\cite{Bowell1989}), which has been adopted in some references,
is known to fail in the case of many TNOs (\cite{Belskaya2008} 2008).
Due to the lack of a better system
we prefer linear methods as a first approximation since TNOs have steep phase curves
which do not deviate from a linear one in the limited phase angle range usually
available for TNO observations. The opposition surge of very small
phase angles ($\alpha \lesssim 0.2\degr$, \cite{Belskaya2008} 2008)
has not been observed for any of our 19 targets due to the lack of observations at
such small phase angles.
All of our \emph{Herschel} and \emph{Spitzer} observations are
limited to the range $1.0\degr < \alpha < 2.1\degr$.

We have used the linear method ($H_{\mathrm{V}}=V-5 \log \left( r \Delta \right)-\beta \alpha$,
where $V$ is the apparent V-magnitude in the Johnson-Coussins or the Bessel V-band, and
other symbols are as in Table~\ref{table_obs}, and
$\beta$ is the linearity coefficient)
to calculate the $H_{\mathrm{V}}$ values from individual V-magnitudes given in the references
(see Table~\ref{table_overview_groundobs}) for 2001 XR$_{254}$, 275809 (2001 QY$_{297}$), 79360 Sila,
88611 Teharonhiawako, 148780 Altjira and 19521 Chaos.
In order to be consistent with most of the $H_{\mathrm{V}}$ values in the literature we have adopted
$\beta=0.14 \pm 0.03$ calculated from \cite{Sheppard2002} (2002).
The effect of slightly different values of $\beta$ or assumptions of its composite value used in
previous works is usually negligible
compared to uncertainties caused by lightcurve variability (an exception is 90568 (2004 GV$_9$)).

For five targets with no other sources available we take the absolute
magnitudes in the R-band ($H_{\mathrm{R}}$ in Table~\ref{table_overview_groundobs}) from the
Minor Planet Center and calculate their standard deviation, since the number of V-band observations
for these targets is very low,
and use the average (V-R) color index for classical TNOs $0.59 \pm 0.15$ (\cite{Hainaut2002} 2002)
to derive $H_{\mathrm{V}}$.
While the MPC is mainly used for astrometry and the magnitudes are considered to be inaccurate
by some works (e.g. \cite{Benecchi2011} 2011, \cite{Romanishin2005} 2005), for the five targets we use the average
of 9 to 16 R-band observations
and adopt the average R-band phase coefficient $\beta_{\mathrm{R}}=0.12$ (calculated from \cite{Belskaya2008} 2008).

The absolute V-magnitudes used as input in our analysis (the ``Corrected $H_{\mathrm{V}}$''
column in Table~\ref{table_overview_groundobs})
take into account additional uncertainties from
known or assumed variability in $H_{\mathrm{V}}$.
The amplitude, the period and the time of zero phase of the lightcurves of
three hot classicals (145452 (2005 RN$_{43}$),
90568 (2004 GV$_9$) and 120347 Salacia) are available in the literature, but even for these
three targets the uncertainty in
the lightcurve period is too large to be used for exact phasing with \emph{Herschel} observations.
The lightcurve amplitude or amplitude limit is known for half of our targets and for them
we add 88\% of the half of the peak-to-peak amplitude ($1\,\sigma$ i.e. 68\% of the values of a sinusoid
are within this range) quadratically to
the uncertainty of $H_{\mathrm{V}}$. According to a study of a sample of 74 TNOs from various dynamical classes
(\cite{Duffard2009}) 70\% of TNOs have a peak-to-peak amplitude $\lesssim0.2$ mag, thus we quadratically add
0.09 mag to the uncertainty of $H_{\mathrm{V}}$ for those targets
in our sample 
for which lightcurve information is not
available.

\begin{table*}
\centering
\caption{Overview of optical auxiliary data. The average absolute V-band (or R-band) magnitudes from
literature are given in Table~\ref{table_overview}, and this table gives the absolute
V-magnitudes with uncertainties which take into account
the lightcurve (L.c.) amplitude $\Delta m_R$
(either in the UBVRI system R-band or Sloan's r'-band). 
The $0\degr$-phase method (see text) tells how the extrapolation to zero phase was done (either in the reference or in this work).
The last column indicates which corrections are significant in contributing to the corrected $H_{\mathrm{V}}$.}
\begin{tabular}{lcccccc}
\hline
\hline
Target                & $H_{\mathrm{V}}$ & $0^{\circ}$-phase method & L.c. $\Delta m_\mathrm{R}$ & L.c. period & Corrected $H_{\mathrm{V}}$ & Corrections \\
                             & ref.             & ([$\beta$]=mag/\degr)      & (mag)             & ($\mathrm{h}$) & (mag) & included  \\
\hline
119951 (2002 KX$_{14}$)      & (a) & phase study             & ...          & ...          & $4.86 \pm 0.10$ & assumed $\Delta m_\mathrm{R}$ \\
(2001 XR$_{254}$)$^*$        & (b) & (1) & ($>$ a few 0.1:s)\tablefootmark{l} & ... & $6.05 \pm 0.15$ & $\Delta m_\mathrm{R}$ \\
275809 (2001 QY$_{297}$)$^*$ & (c,d,e) & (1) & 0.49--0.66\tablefootmark{c} & ...  & $5.86 \pm 0.31$ & $\Delta m_\mathrm{R}$ \\
(2001 RZ$_{143}$)$^*$        & (f) & $\beta=0.16$\tablefootmark{f}  & ...               & ...                           & $6.69 \pm 0.13$ & assumed $\Delta m_\mathrm{R}$ \\ 
(2002 GV$_{31}$)	     & (g) & (2) & ...               & ...      & $6.1  \pm 0.6$ & \ldots \\
79360 Sila$^*$ & (h,i,j,k) & (1) & $<$ 0.08\tablefootmark{u} & ... & $5.56 \pm 0.04$ & $\Delta m_\mathrm{R}$ \\ 
88611 Teharonhiawako$^*$  & (l) & (1) & 0.2\tablefootmark{v}  & $4.7526 \pm 0.0007$\tablefootmark{v} & $6.00 \pm 0.13$ & $\Delta m_\mathrm{R}$ \\
120181 (2003 UR$_{292}$)     & (g) & $\beta_\mathrm{R}=0.14$\tablefootmark{t}                    &     ...    & ...      & $7.4  \pm 0.4$ & \ldots \\
(2005 EF$_{298}$)            & (g) & (2)  & ...       & ...      & $6.4  \pm 0.5$ & \ldots \\
\hline
138537 (2000 OK$_{67}$)      & (m,n) & $\beta=0.15$\tablefootmark{m} / $0.14$\tablefootmark{n} &... & ... & $6.47 \pm 0.13$ & assumed $\Delta m_\mathrm{R}$ \\
148780 Altjira$^*$        & (o,p,d) & (1) & $<$ 0.3\tablefootmark{p} & ... & $6.44 \pm 0.14$ & $\Delta m_\mathrm{R}$ \\ 
(2002 KW$_{14}$) & (q) & (1) & 0.21 / 0.26\tablefootmark{c} & ... & $5.88 \pm 0.11$ & $\Delta m_\mathrm{R}$ \\
(2001 KA$_{77}$)	     & (o) & (1)  & ...               & ...                            & $5.64 \pm 0.12$ & assumed $\Delta m_\mathrm{R}$ \\
 19521 Chaos & (h,i,n,r) & (1) & $<$ 0.10\tablefootmark{u} & ... & $5.00 \pm 0.06$ & $\Delta m_\mathrm{R}$ \\
78799 (2002 XW$_{93}$)       & (g) & (2) & ...               & ...                         & $5.4  \pm 0.7$  & \ldots \\
(2002 MS$_4$)	             & (g) & (2) & ...               & ...                         & $4.0  \pm 0.6$  & \ldots \\
145452 (2005 RN$_{43}$)      & (q) & (1) & $0.04 \pm 0.01^p$ & $5.62 / 7.32$\tablefootmark{x} & $3.89 \pm 0.05$ & $\Delta m_\mathrm{R}$ \\
90568 (2004 GV$_9$) & (s) & $\beta=0.18 \pm 0.06$\tablefootmark{p} & $0.16 \pm 0.03$\tablefootmark{w} & $5.86 \pm 0.03$\tablefootmark{w} & $4.23 \pm 0.10$ & $\beta$, $\Delta m_\mathrm{R}$ \\
120347 Salacia$^*$    & (q) & (1) & $0.03 \pm 0.01$\tablefootmark{x} & $6.09 / 8.1$\tablefootmark{x} & $4.24 \pm 0.04$ & $\Delta m_\mathrm{R}$ \\
\hline
\end{tabular}
\label{table_overview_groundobs}
\tablefoot{* denotes a known binary system (\cite{Noll2008}).
(1) V-band linear phase coefficient $\beta=0.14 \pm 0.03$ (\cite{Sheppard2002} 2002,
$\beta$ calculated from values therein).
(2) R-band linear phase coefficient $\beta_\mathrm{R}=0.12$ (Average from \cite{Belskaya2008} 2008). 
{\bf References.}
\tablefoottext{a}{\cite{Rabinowitz2007}.}
\tablefoottext{b}{\cite{Grundy2009}.}
\tablefoottext{c}{\cite{Thirouin2012} (2012).}
\tablefoottext{d}{\cite{Grundy2011}.}
\tablefoottext{e}{\cite{Doressoundiram2007}.}
\tablefoottext{f}{\cite{SantosSanz2009} (2009).}
\tablefoottext{g}{IAU Minor Planet Center / List of Transneptunian Objects at
\url{http://www.minorplanetcenter.net/iau/lists/TNOs.html}, accessed June 2011.}
\tablefoottext{h}{\cite{Bohnhardt2001}.}
\tablefoottext{i}{\cite{Davies2000}.}
\tablefoottext{j}{\cite{Barucci2000}.}
\tablefoottext{k}{\cite{Grundy2012}.}
\tablefoottext{l}{\cite{Benecchi2009}.}
\tablefoottext{m}{\cite{Benecchi2011} (2011).}
\tablefoottext{n}{\cite{Doressoundiram2002}.}
\tablefoottext{o}{\cite{Doressoundiram2005}.}
\tablefoottext{p}{\cite{Sheppard2007}.}
\tablefoottext{q}{Perna et al., {\it in prep}.}
\tablefoottext{r}{\cite{Tegler2000}.}
\tablefoottext{s}{\cite{DeMeo2009}.}
\tablefoottext{t}{Calculated from MPC data from 16 observations.}
\tablefoottext{u}{\cite{Sheppard2002} (2002).}
\tablefoottext{v}{\cite{Osip2003}, lightcurve caused by secondary component, amplitude calculated for the combined system.}
\tablefoottext{w}{\cite{Dotto2008}.}
\tablefoottext{x}{\cite{Thirouin2010} (2010).}
}
\end{table*}

\section{Thermal modeling}
\label{model}
The combination of observations from thermal-infrared and optical wavelengths
allows us to estimate various physical properties via thermal modeling.
For a given temperature distribution
the disk-integrated thermal emission $F$ observed at wavelength $\lambda$ is 
\begin{equation}
F(\lambda) = \frac{\epsilon\left(\lambda\right)}{\Delta^2}
\int_{S} B \left( \lambda, T\left( S \right)\right) \;d\textbf{S} \cdot \textbf{u}, 
\end{equation}
where $\epsilon$ is the emissivity, $\Delta$ the observer-target distance,
$B(\lambda, T)$ Planck's radiation law for black bodies, $T \left( S \right)$
the temperature distribution on the surface $S$ and $\textbf{u}$ the
unit directional vector toward the observer from the surface element $d\textbf{S}$.
The temperature distribution of an airless body depends on physical parameters
such as diameter, albedo, thermal inertia, and surface roughness. 

There are three basic types of models to predict the emission of an
asteroid with a given size and albedo assuming an equilibrium
between insolation and re-emitted thermal radiation:
the Standard Thermal Model (STM),
the Fast-rotating Isothermal Latitude thermal model (\cite{Veeder1989}), and the thermophysical models
(starting from \cite{Matson1971}, e.g. \cite{Spencer1989,Lagerros1996}).
While originally developed for asteroids in the mid-IR wavelengths these models are applicable for
TNOs, whose thermal peak is in the far-IR. 

The STM (cf. \cite{Lebofsky1986} and
references therein) assumes a smooth, spherical asteroid, which is not
rotating and/or has zero thermal inertia, and is observed at zero phase angle.
The subsolar temperature $T_{\mathrm{SS}}$ is
\begin{equation}
\label{Tss}
T_{\mathrm{SS}} = \left[\frac{ \left(1-A\right)S_{\sun}}{\epsilon \eta \sigma r^2} \right]^{\frac{1}{4}},
\end{equation}
where $A$ is the Bond albedo, $S_{\sun}$ is the solar constant, $\eta$ is the beaming factor,
$\sigma$ is the Stefan-Boltzmann constant and $r$ is the heliocentric distance. In the STM
$\epsilon$ does not depend on wavelength. The beaming factor $\eta$ adjusts the subsolar temperature.
The canonical value $\eta=0.756$ is based on calibrations using the largest few main belt asteroids. 
The STM assumes an average linear infrared phase coefficient of 0.01 mag/degree based on
observations of main belt asteroids. 

In this work we use the Near-Earth Asteroid Thermal Model NEATM (\cite{Harris1998}).
The difference to the STM is that in the NEATM $\eta$ is fitted
with the data instead of using a single canonical value. For rough surfaces $\eta$ takes
into account the fact that points on the surface $S$ radiate
their heat preferentially in the sunward direction. High values ($\eta>1$) lead to
a reduction of the model surface temperature, mimicking the effect of high thermal inertia,
whereas lower values are a result of surface roughness.
Furthermore, 
the phase angle $\alpha$ is
taken into account by calculating the thermal flux an observer would detect from the
illuminated part of $S$ assuming a Lambertian emission model and no emission from the
non-illuminated side.

Whenever data quality permits we treat $\eta$ as a free fitting parameter.
However, in some cases of poor data quality
this method leads to $\eta$ values which are too high or too low and therefore unphysical.
In these cases we fix it to a canonical value of $\eta = 1.20 \pm 0.35$ derived by
\cite{Stansberry2008} (2008) from \emph{Spitzer} observations of TNOs. 
The physical range of $\eta$ values is determined by using NEATM 
as explained in \cite{Mommert2012} (2012) to be $0.6 \leq \eta \leq 2.6$.

Throughout this work we assume the surface emissivity $\epsilon=0.9$, which is
based on laboratory measurements of silicate powder up to a wavelength of
$22\ \mathrm{\mu m}$ (\cite{Hovis1966}) and a usual
approximation for small bodies in the Solar System. At far-IR wavelengths
the emissivity of asteroids may be decreasing as a function of wavelength
(\cite{Muller1998}) from 24 to
$160\ \mathrm{\mu m}$, but the amount depends on individual target properties. Thus,
a constant value is assumed for simplicity. The variation of emissivity of
icy surfaces as a function of wavelength could in principle provide hints
about surface composition. {H$_2$O} ice has an emissivity close to one
with small variations (\cite{Schmitt1998}) whereas other ices may show a stronger
wavelength dependence (e.g. \cite{Stansberry1996}). Near-IR spectroscopic surface studies
have been done for five of our targets but none of them show a reliable detection
of ices, even though many dynamically hot classicals are known to have
ice signatures in their spectra (\cite{Barucci2011} 2011).

For the Bond albedo
we assume that $A\approx A_V$ since the V-band is close to the peak of the
solar spectrum. The underlying assumption is that the Bond albedo is not
strongly varying across the relevant solar spectral range of reflected light.
The Bond albedo is connected to the geometric albedo via $A_V=p_Vq$, where
$q$ is the phase integral and $p_V$ the geometric albedo in V-band. Instead
of the canonical value of $q=0.39$
(\cite{Bowell1989}) we have adopted $q=0.336 p_V+0.479$
(\cite{Brucker2009} 2009), which they used for a \emph{Spitzer} study of classical TNOs.
From the definition of the absolute magnitude of asteroids
we have:
\begin{equation}
p_V S_\mathrm{proj}=\pi a^2 \times 10^{\frac{2}{5} \left( m_\mathrm{\sun} - H_V \right)},
\label{pA}
\end{equation}
where $S_\mathrm{proj}$ is the area projected
toward the observer, $a$ is the distance of one astronomical unit, $m_\mathrm{\sun}$
is the apparent V-magnitude of the Sun, and $H_V$ is the absolute
V-magnitude of the asteroid.
We use $m_\mathrm{\sun} = \left( -26.76 \pm 0.02 \right)$ mag (\cite{Bessell1998,Hayes1985}).
An error of 0.02 mag in $H_V$ means a relative error of $1.8\%$ in the product (\ref{pA}).

We find the free parameters $p_V$, $D=\sqrt{\frac{4 S_\mathrm{proj}}{\pi}}$ and $\eta$
in a weighted least-squares sense by minimizing the cost function
\begin{equation}
 \chi_{\nu}^2=\frac{1}{\nu} \sum_{i=1}^{N} \frac{\left[ F \left( \lambda_i \right) -
 F_{\mathrm{model}} \left( \lambda_i \right)\right]^2}{\sigma_i^2},
\label{chi2}
\end{equation}
where $\nu$ is the number of degrees of freedom,
$N$ is the number of data points in the far-infrared wavelengths,
$F \left( \lambda_i \right)$ is the observed flux density at wavelength $\lambda_i$ with uncertainty
$\sigma_i$, and $F_{\mathrm{model}}$ is the modeled emission spectrum.
When only upper flux density
limits are available, they are treated as having zero flux density with a $1\,\sigma$ uncertainty
equal to the upper limit flux density uncertainty ($0\pm\sigma$). The cost function
(\ref{chi2}) does not follow the $\chi^2$ statistical distribution since in a non-linear
fit the residuals are not normally distributed even if the flux densities had normally
distributed uncertainties.

\subsection{Error estimates}
The error estimates of the geometric albedo, the diameter and the beaming parameter are determined by a Monte
Carlo method described in \cite{Mueller2011}. We generate 500 sets of synthetic
flux densities normally distributed around the observed flux densities with the same standard
deviations as the observations. Similarly, a set of normally distributed $H_{\mathrm{V}}$ values
is generated. For those targets whose $\chi^2 >>1$ we do a rescaling of the errorbars
of the flux densities before applying the Monte Carlo error estimation as described in
\cite{SantosSanz2012} (2012) and illustrated in \cite{Mommert2012} (2012).

The NEATM model gives us the effective diameter of a spherical target.
70-80\% of TNOs are known to be MacLaurin spheroids with an axial ratio of 1.15
(\cite{Duffard2009}, \cite{Thirouin2010} 2010). When the projected surface has the shape of an ellipse instead of a
circle,  then this ellipse will emit more
flux than the corresponding circular disk which has the same surface
because the Sun is seen at higher elevations from a larger portion of the ellipsoid
than on the sphere. Therefore, the NEATM diameters may be slightly overestimated.
Based on studies of model accuracy (e.g. \cite{Harris2006})
we adopt uncertainties of 5\% in the diameter estimates and 10\% in the $p_{\mathrm{V}}$
estimates to account for systematic model errors 
when NEATM is applied
at small phase angles.

\section{Results of individual targets}
\label{resultsection}
In this Section we give the results of model fits using the NEATM
to determine the area-equivalent diameters (see Section~\ref{model}) as well
as geometric albedos and beaming factors. 
We note that our observations did not spatially resolve binary systems, we therefore
find area-equivalent diameters of the entire system rather than component diameters;
this will be further discussed in Section~\ref{binaries}.

In cases where also
\emph{Spitzer}/MIPS data are available for a target we determine the free parameters for both PACS only
and the combined data sets. The solutions are given in Table~\ref{table_results}.
A floating-$\eta$ solution is only
adopted if 
its $\chi^2$ is not much greater than unity. The exact limit
depends on the number of data points. For $N=5$ this limit is $\chi^2 \lesssim 1.7$.
For the PACS-only data set we may adopt
floating-$\eta$ solutions only if there are no upper-limit data points.
138537 (2000 OK$_{67}$) has only upper limits from PACS,
therefore only the solution using combined data is shown.
2001 RZ$_{143}$ has five data points, three of which are upper limits. The PACS flux density at $70\ \mathrm{\mu m}$
is approximately a factor of three higher than the MIPS upper limit (see
Tables~\ref{table_Pfluxes} and \ref{table_Spitzerobs}).
For this target we adopt the
model solution determined without the MIPS $70\ \mathrm{\mu m}$ channel. 
The best solution for each target is shown in Fig.~\ref{fits_lot1}. 

The radiometric diameters determined with data from the two instruments are on the average close to
the corresponding results using PACS data alone, but there are some significant differences as well,
most notably 275809 (2001 QY$_{297}$) and 2002 KW$_{14}$. The former has a
PACS-only solution, which is within the error bars of the green and red channel data points,
but above the PACS blue channel data point.
When the two upper limits from MIPS are added in the analysis the model solution is at a
lower flux level below the PACS green channel data point but compatible with
the other data (see Fig.~\ref{fits_lot1}).
2002 KW$_{14}$ has upper limits in the PACS green and red channels as well as in the MIPS
$24\ \mathrm{\mu m}$ channel. Without this upper limit in the shortest wavelength the PACS-only
solution is at higher flux levels in short wavelengths and gives a lower flux at long wavelengths.

When choosing the preferred solution we are comparing two fits with different numbers of
data points used, therefore in this comparison we calculate $\chi^2$ for the PACS-only
solution using the same data points as for the combined solution taking into account
the different observing geometries during \emph{Herschel} and \emph{Spitzer} observations.
In all cases where MIPS data is available the solution based on the combined data from the two instruments is the
preferred one.

2002 GV$_{31}$ is
      the only non-detection by both PACS and MIPS in our sample of 19 targets. The astrometric $3\,\sigma$
      uncertainty at the time of the PACS observation was $<4\arcsec$ (semimajor axis of the
      confidence ellipsoid\footnote{Asteroids Dynamic Site by A. Milani, Z. Knezevic, O. Arratia et al.,
      \url{http://hamilton.dm.unipi.it/astdys/}, accessed August 2011, calculations based on the
      OrbFit software.}), which is well within the
      high-coverage area of our maps.

The error bars from the Monte Carlo error estimation method may sometimes be too
optimistic compared to the accuracy of 
optical data
and the model uncertainty of NEATM (see Section~\ref{model}). 
We check that the uncertainty of geometric albedo is not
better than the uncertainty implied by the optical constraint (Eq.~\ref{pA}) due to
uncertainties in $H_{\mathrm{V}}$ and $m_{\sun}$. The lower $p_{\mathrm{V}}$ uncertainty
of four targets is limited by this (see Table~\ref{table_results}).

\begin{table*}
\centering
\caption{Solutions for radiometric diameters and geometric albedos (see text
for explanations). For binary systems $D$ is the area-equivalent system diameter. In case of two
solutions the preferred one is based on data from both PACS and MIPS instruments.}
\begin{tabular}{llcccc}
\hline
Target & Instruments & $D$    & Bin- & $p_\mathrm{V}$\tablefootmark{a} & $\eta$ \\
       &             & (km) & ary? &       &        \\
\hline
\hline
119951 (2002 KX$_{14}$) & PACS          & $485_{-93}^{+83}$     & no  & $0.086_{-0.023}^{+0.042}$       & $2.07_{-0.82}^{+0.88}$ \\
                    & {\bf PACS, MIPS} & ${\bf 455 \pm 27}$ &     & ${\bf 0.097_{-0.013}^{+0.014}}$ & ${\bf 1.79_{-0.15}^{+0.16}}$ \\
                             &                        &                    &     &                                 &  \\
(2001 XR$_{254}$)            & PACS          & $200_{-63}^{+{49}}$ & yes & $0.17_{-0.05}^{+0.19}$ & $1.2 \pm 0.35$ (fixed) \\
                             &                        &                    &     &                                 &  \\
275809   (2001 QY$_{297}$) & {PACS} & ${278_{-55}^{+45}}$ & yes & ${0.104_{-0.050}^{+0.094}}$ & ${1.2 \pm 0.35}$ {(fixed)} \\
                        & {\bf PACS, MIPS} & ${\bf 200_{-59}^{+62}}$ & & ${\bf 0.20_{-0.11}^{+0.25}}$ & ${\bf 1.2 \pm 0.35}$ {\bf (fixed)} \\
                             &                        &                    &     &                                 &  \\
(2001 RZ$_{143}$)      & PACS                         & $168_{-52}^{+47}$  & yes & $0.13_{-0.05}^{+0.15}$ & $1.2 \pm 0.35$ (fixed) \\
                  & {\bf PACS, MIPS\tablefootmark{b}} & ${\bf 140_{-33}^{+39}}$ & & ${\bf 0.191_{-0.045}^{+0.066}}$ & ${\bf 0.75_{-0.19}^{+0.23}}$    \\
                             &                        &                    &     &                                 &  \\
(2002 GV$_{31}$)       & PACS                         & $<130$             & no  & $>0.22$          & $1.2 \pm 0.35$ (fixed) \\
                             &                        &                    &     &                                 &  \\
79360 Sila       & PACS                  & $333_{-47}^{+38}$ & yes & $0.095_{-0.020}^{+0.033}$ & $1.2 \pm 0.35$ (fixed) \\
                             & {\bf PACS, MIPS}       & ${\bf 343 \pm 42}$  &   & ${\bf 0.090_{-0.017}^{+0.027}}$ & ${\bf 1.36_{-0.19}^{+0.21}}$  \\
                             &                        &                    &     &                                 &  \\
88611      Teharonhiawako & PACS                    & $234_{-39}^{+37}$ & yes & $0.129_{-0.036}^{+0.062}$ & $1.2 \pm 0.35$ (fixed) \\
     & {\bf PACS, MIPS} & ${\bf 177_{-44}^{+46}}$ & & ${\bf 0.22_{-0.08}^{+0.14}}$ & ${\bf 0.86_{-0.29}^{+0.37}}$ \\
                             &                        &                    &     &                                 &  \\
120181 (2003 UR$_{292}$)      & PACS                  & $104_{-24}^{+22}$   & no & $0.16_{-0.08}^{+0.19}$ & $1.2 \pm 0.35$ (fixed) \\
                             &                        &                    &     &                                 &  \\
(2005 EF$_{298}$)   & PACS                            & $174_{-32}^{+27}$   & no & $0.16_{-0.07}^{+0.13}$ & $1.2 \pm 0.35$ (fixed) \\
\hline
138537  (2000 OK$_{67}$)   & PACS, MIPS               & $151_{-37}^{+31}$  & no  & $0.20_{-0.08}^{+0.21}$ & $1.2 \pm 0.35$ (fixed) \\
                             &                        &                    &     &                                 &  \\
148780      Altjira & PACS                         & $313_{-48}^{+50}$ & yes & $0.048_{-0.015}^{+ 0.022}$ & $1.2 \pm 0.35$ (fixed) \\
         & {\bf PACS, MIPS} & $\bf 257_{-92}^{+90}$ & & $\bf 0.071_{-0.021}^{+0.049}$ & $\bf 1.46 \pm 0.41$ \\
                             &                        &                    &     &                                 &  \\
(2002 KW$_{14}$)    & PACS                            & $181_{-38}^{+36}$   & no & $0.24_{-0.07}^{+0.14}$ & $1.2 \pm 0.35$ (fixed) \\
                    & {\bf PACS, MIPS}                & $\bf {319_{-81}^{+74}}$ & & ${\bf 0.08_{-0.05}^{+0.14}}$ & ${\bf 2.50_{-1.64}^{+0.14}}$  \\
                             &                        &                    &     &                                 &  \\
(2001 KA$_{77}$)     & PACS                           & $252_{-57}^{+49}$ & no & $0.155_{-0.046}^{+0.091}$ & $1.2 \pm 0.35$ (fixed) \\
                     & {\bf PACS, MIPS}               & ${\bf 310_{-60}^{+170}}$ & & ${\bf 0.099_{-0.056}^{+0.052}}$ & ${\bf 2.52_{-0.83}^{+0.18}}$    \\
                             &                        &                    &     &                                 &  \\
19521        Chaos & PACS         & $600_{-150}^{+140}$  & no  & $0.050_{-0.019}^{+0.040}$ & $2.2_{-1.0}^{+1.6}$         \\
& {\bf PACS, MIPS} & $\bf 600_{-130}^{+140}$ & & $\bf 0.050_{-0.016}^{+0.030}$ & $\bf 2.2_{-1.1}^{+1.2}$   \\
                             &                        &                    &     &                                 &  \\
78799    (2002 XW$_{93}$) & PACS           & $565_{-73}^{+71}$ & no & $0.038_{-0.025}^{+0.043}$ & $0.79_{-0.24}^{+0.27}$ \\ 
                             &                        &                    &     &                                 &  \\
(2002 MS$_4$)      & PACS                           & $988_{-105}^{+96}$ & no & $0.046_{-0.026}^{+0.042}$ & $1.2 \pm 0.35$ (fixed) \\
                   & {\bf PACS, MIPS}              & ${\bf 934 \pm 47}$\tablefootmark{c} & & ${\bf 0.051_{-0.022}^{+0.036}}$ & ${\bf 1.06 \pm 0.06}$ \\

                             &                        &                    &     &                                 &  \\
145452  (2005 RN$_{43}$)  & PACS                     & $679_{-73}^{+55}$ & no & $0.107_{-0.018}^{+ 0.029}$ & $1.2 \pm 0.35$ (fixed) \\
                             &                        &                    &     &                                 &  \\
90568  (2004 GV$_9$)    & PACS                        & $670_{-180}^{+210}$ & no  & $0.08_{-0.06}^{+ 0.15}$ & $1.9_{-0.8}^{+1.5}$ \\
                        & {\bf PACS, MIPS} & ${\bf 680 \pm 34}$\tablefootmark{c} & & ${\bf 0.0770_{-0.0077}^{+0.0084}}$ & ${\bf 1.93_{-0.07}^{+0.09}}$  \\
                             &                        &                    &     &                                 &  \\
120347      Salacia & PACS                            & $994_{-74}^{+88}$  & yes & $0.0361_{-0.0055}^{+0.0059}$ & $1.47_{-0.25}^{+0.29}$    \\
      & {\bf PACS, MIPS} & $\bf 901 \pm 45$\tablefootmark{c} & & $\bf 0.0439 \pm 0.0044$\tablefootmark{d} & $\bf 1.156 \pm 0.031$ \\
\hline
\end{tabular}
\label{table_results}
\tablefoot{
\tablefoottext{a}{Lower uncertainty limited by the uncertainty of $H_{\mathrm{V}}$ for 275809 / PACS-MIPS, 2005 EF$_{298}$, 78799, and
2002 MS$_4$ / both solutions.}
\tablefoottext{b}{MIPS $70\ \mathrm{\mu m}$ channel excluded.}
\tablefoottext{c}{Error estimate limited by the adopted diameter uncertainty of 5\% of the NEATM model.}
\tablefoottext{d}{Error estimate limited by the adopted $p_{\mathrm{V}}$ uncertainty of 10\% of the NEATM model.}}
\end{table*}

\begin{figure*}
   \includegraphics[width=17cm]{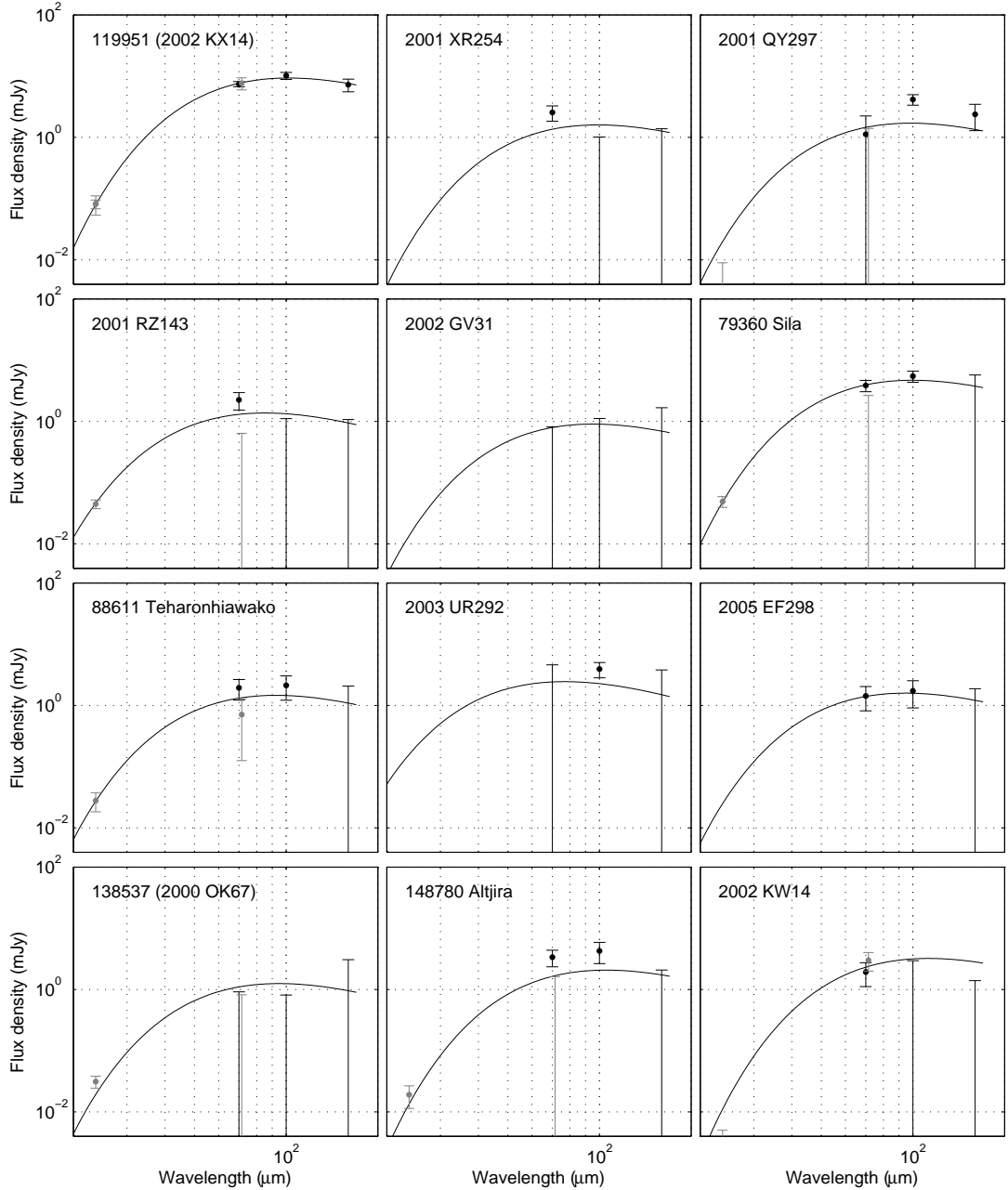}
   \caption{Adopted model solutions from Table~\ref{table_results}. The black data points are from PACS
(70, 100 and $160\ \mathrm{\mu m}$) and the gray points are from MIPS
(24 and $71\ \mathrm{\mu m}$) normalized to the geometry of \emph{Herschel}
observations by calculating the NEATM solution (for given $D$, $p_{\mathrm{V}}$ and $\eta$) at both the epochs
of the \emph{Herschel} and \emph{Spitzer} observations and using their ratio as a correction factor.}
   \label{fits_lot1}
\end{figure*}
\setcounter{figure}{2} 
\begin{figure*}
   \includegraphics[width=17cm]{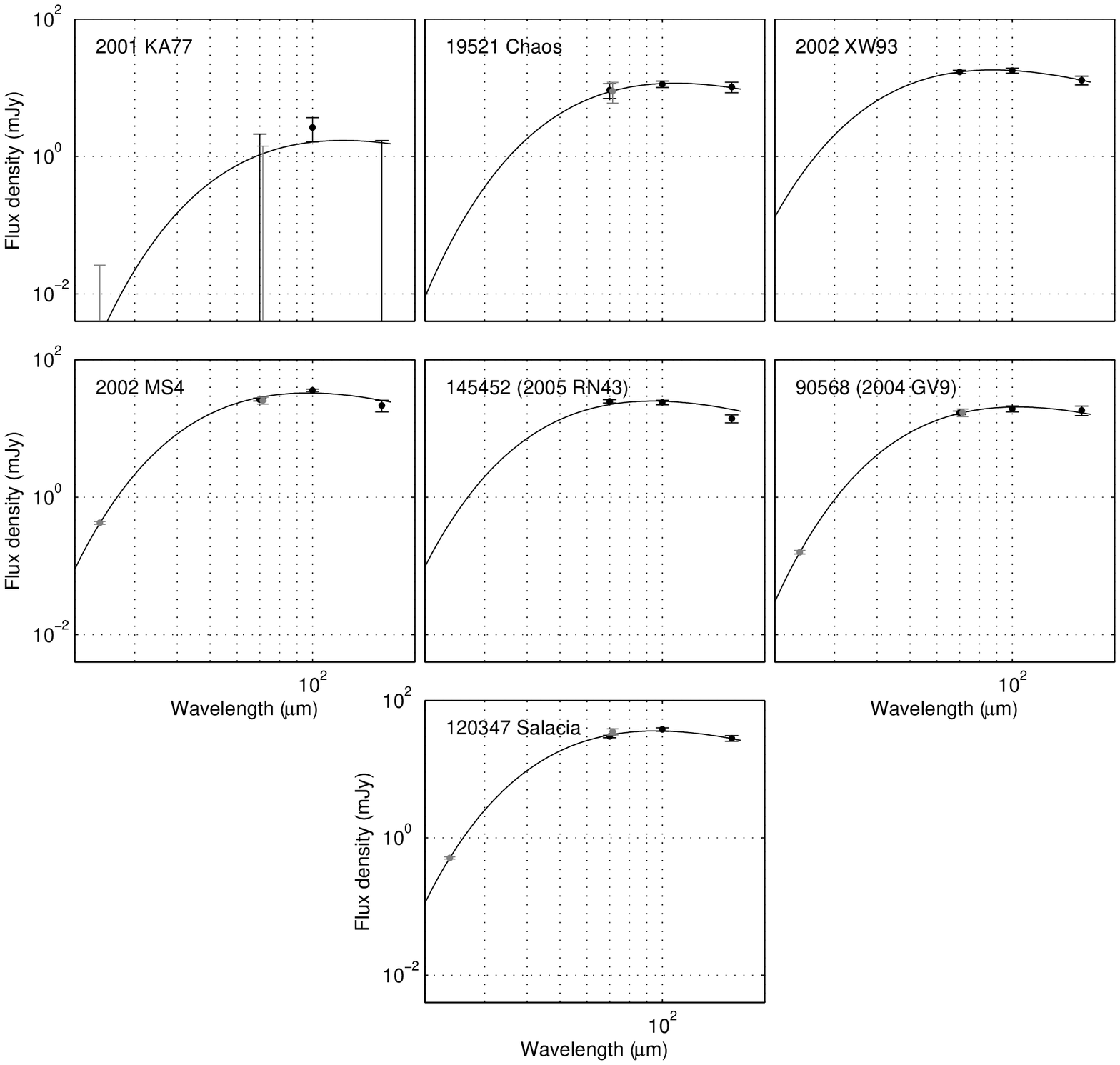} 
   \caption{continued.}
   \vspace{15mm}
   \label{fits_lot2}
\end{figure*}

\section{Discussion}
\label{discussions}
Our new size and geometric albedo estimates improve the accuracy of previous estimates of
almost all the targets with existing results or limits (Table~\ref{collection}).
Two thirds
of our targets have higher albedos (Table~\ref{table_results}) than the $0.08$ used in the planning
of these observations, which has lead to the lower than expected SNRs
and several upper limit flux densities (Table~\ref{table_Pfluxes}).
Previous results from \emph{Spitzer} are generally compatible with our new estimates.
However, two targets are significantly different: 119951 (2002 KX$_{14}$)
and 2001 KA$_{77}$, whose solutions are compatible with the optical constraint (Eq.~\ref{pA}),
but the new estimate has a large diameter and low albedo (target 119951) instead of a small
diameter and high albedo,
or vice versa (target 2001 KA$_{77}$).
It can be noted that 
there is a significant difference in the re-processed
\emph{Spitzer} flux densities        
 at $70\ \mathrm{\mu m}$ 
compared to
the previously published 
values (see Table~\ref{table_Spitzerobs}),
which together with the PACS $100\ \mathrm{\mu m}$ data can explain the significant change in the diameter and
geometric albedo estimates. 
Our new estimates for 148780 Altjira differ from the previous
upper and lower limits based on \emph{Spitzer} data alone (Table~\ref{collection}). This change can be
explained by the addition of the $70\ \mathrm{\mu m}$ and $100\ \mathrm{\mu m}$ PACS data points (Table~\ref{table_Pfluxes})
to the earlier MIPS $24\ \mathrm{\mu m}$ value and the MIPS $70\ \mathrm{\mu m}$ upper flux density limit (Table~\ref{table_Spitzerobs}).

\begin{table*}
\centering
\caption{Adopted physical properties in comparison with previous works. Binary systems are indicated by 'B'.
Column $\lambda_{\mathrm{detect}}$
lists the wavelengths used in the reference for radiometric diameters. If no radiometric result is available
then binary system mass is used to give the diameter of the primary component assuming equal albedos. The second part of
this table lists all other classical TNOs with radiometric or other reliable size estimates. The size of
50000 Quaoar is derived from both \emph{Spitzer} and from direct imaging by \emph{Hubble}. All targets in the lower part
are dynamically hot.}
\begin{tabular}{ll|cl|lcll}
\hline
         & & \multicolumn{2}{c}{This work} & \multicolumn{3}{c}{Previous work} \\
  Target & &   D (km) & p$_V$ & $\lambda_{\mathrm{detect}}$ ($\mu m$) & D (km) &  p$_V$ & Reference \\
\hline
119951  (2002 KX$_{14}$)  &   & $455 \pm 27$    & $0.097_{-0.013}^{+0.014}$  & 24, 71  & $180_{-38}^{+50}$ & $0.60_{-0.23}^{+0.36}$ & \cite{Brucker2009} (2009) \\
(2001 XR$_{254}$)       & B & $200_{-63}^{+49}$ & $0.17_{-0.05}^{+0.19}$ & (binary) & 130--208\tablefootmark{a} & 0.09--0.23 & \cite{Grundy2011} \\
275809  (2001 QY$_{297}$) & B & $200_{-59}^{+62}$ & $0.20_{-0.11}^{+0.25}$ & (binary) & 128--200\tablefootmark{a} & 0.13--0.32 & \cite{Grundy2011}   \\
(2001 RZ$_{143}$)         & B &  $140_{-33}^{+39}$ &  $0.191_{-0.045}^{+0.066}$ & 24      & $<160$    & $>0.23$ & \cite{Brucker2009} (2009) \\
79360 Sila         & B &  $343 \pm 42$ & $0.090_{-0.017}^{+0.027}$  & 70, 160 & 250--420  & 0.06-0.14 & \cite{Muller2010} (2010) \\
88611 Teharonhiawako & B & $177_{-44}^{+46}$ & $0.22_{-0.08}^{+0.14}$ & (binary) & 114--180\tablefootmark{a} &  0.13--0.32 & \cite{Grundy2011} \\
138537 (2000 OK$_{67}$)   &   & $151_{-37}^{+31}$  & $0.20_{-0.08}^{+0.21}$     & 24       & $<160$     & $>0.16$  & \cite{Brucker2009} (2009) \\
148780 Altjira   & B & $257_{-92}^{+90}$  & $0.071_{-0.021}^{+0.049}$  & 24 & $<200$ & $>0.10$ & \cite{Brucker2009} (2009) \\
                         &   &                 &                        & (binary) & 128--200\tablefootmark{a} & 0.06--0.14 & \cite{Grundy2011}   \\
(2002 KW$_{14}$)          &   & $319_{-81}^{+74}$  & $0.08_{-0.05}^{+0.14}$     & 71       & $<360$   &   $>0.05$ & \cite{Brucker2009} (2009) \\
(2001 KA$_{77}$)          &   & $310_{-60}^{+170}$  & $0.099_{-0.056}^{+0.052}$  & 24, 71   & $634_{-92}^{+134}$ & $0.025_{-0.008}^{+0.010}$ & \cite{Brucker2009} (2009) \\
19521    Chaos           &   & $600_{-130}^{+140}$  & $0.050_{-0.016}^{+0.030}$  & 1200 & $<742$ &  $>0.033$ & \cite{Altenhoff2004} \\
(2002 MS$_4$)            &   & $934 \pm 47$  & $0.051_{-0.022}^{+0.036}$ & 24, 71  & $730 \pm 120$ & $0.073_{-0.03}^{+0.06}$ & \cite{Brucker2009} (2009) \\
90568 (2004 GV$_9$)   & & $680 \pm 34$ & $0.0770_{-0.0077}^{+0.0084}$ & 24, 71 & $684_{-74}^{+68}$ & $0.073_{-0.03}^{+0.05}$ & \cite{Brucker2009} (2009) \\
120347      Salacia      & B & $901 \pm 45$  & $0.0439 \pm 0.0044$ & (binary) & 720--1140\tablefootmark{a} & 0.01--0.03 & \cite{Grundy2011} \\
                         &   &                 &                    & 24, 71   & $954 \pm 80$    & $0.0357_{-0.0056}^{+0.0072}$ & \cite{Stansberry2012} \\
\hline
(2001 QD$_{298}$)        &    & \multicolumn{2}{c|}{\ldots} & 24, 71 & $150_{-40}^{+50}$  & $0.18_{-0.08}^{+0.17}$ & \cite{Brucker2009} (2009) \\ 
(1996 TS$_{66}$)         &    & \multicolumn{2}{c|}{\ldots} & 24, 71 & $190_{-40}^{+50}$  & $0.12_{-0.05}^{+0.07}$ &  \cite{Brucker2009} (2009) \\ 
50000 Quaoar            & B  & \multicolumn{2}{c|}{\ldots} & 24, 71, (direct) & $890 \pm 70$\tablefootmark{a} & $0.18 \pm 0.04$ & \cite{Fraser2010b} (2010) \\ 
(2002 GJ$_{32}$)         &    & \multicolumn{2}{c|}{\ldots} & 24, 71 & $220_{-70}^{+90}$  & $0.12_{-0.06}^{+0.14}$ &  \cite{Brucker2009} (2009) \\ 
20000 Varuna            &    & \multicolumn{2}{c|}{\ldots} & 71     & $710_{-130}^{+180}$  & $0.09_{-0.03}^{+0.04}$ & \cite{Brucker2009} (2009) \\ 
55637   (2002 UX$_{25}$) & B  & \multicolumn{2}{c|}{\ldots} & 24, 71 & $680_{-110}^{+120}$  & $0.12_{-0.03}^{+0.05}$ & \cite{Stansberry2008} (2008) \\ 
55636 (2002 TX$_{300}$)  &    & \multicolumn{2}{c|}{\ldots} & (occultation) & $286 \pm 10$  & $0.88_{-0.06}^{+0.15}$ & \cite{Elliot2010} \\ 
55565 (2002 AW$_{197}$)  &    & \multicolumn{2}{c|}{\ldots} & 24, 71 & $740 \pm 100$        & $0.12_{-0.03}^{+0.04}$ & \cite{Brucker2009} (2009) \\ 
\hline
\end{tabular}
\label{collection}
\tablefoot{
\tablefoottext{a}{Diameter of the primary component.}
}
\end{table*}

\begin{figure}
    \centering
   \includegraphics[width=10cm]{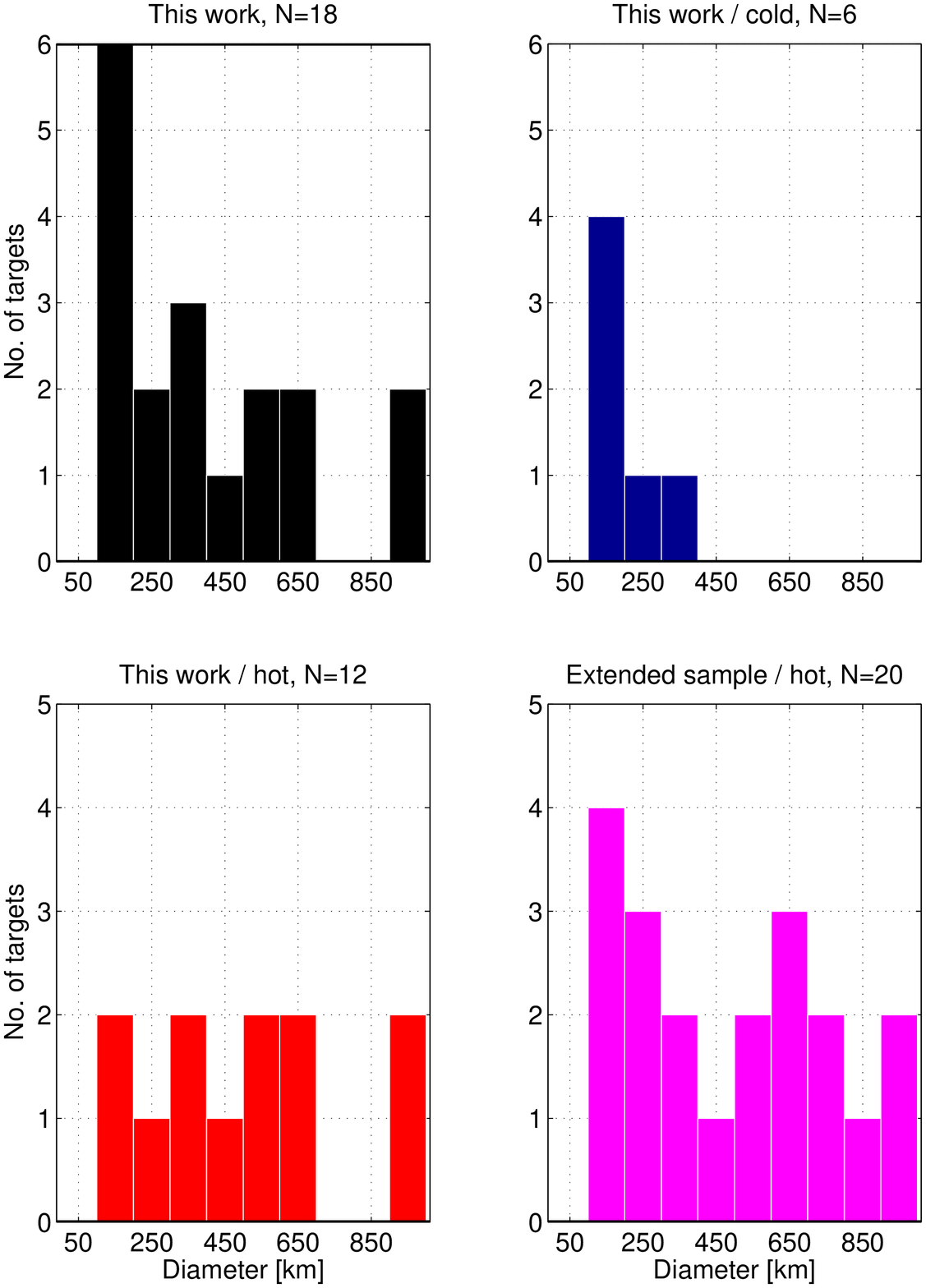}
   \caption{Distribution of diameters from this work (upper left), the cold classicals
of this work (upper right), the hot classicals of this work (lower left), and all hot classicals
including literature results from Table~\ref{collection} (lower right). The last plot includes only dynamically
hot classicals. The bin size is 100 km.}
   \label{diameter_histo}
\end{figure}

\begin{figure}
   \centering
   \includegraphics[width=10cm]{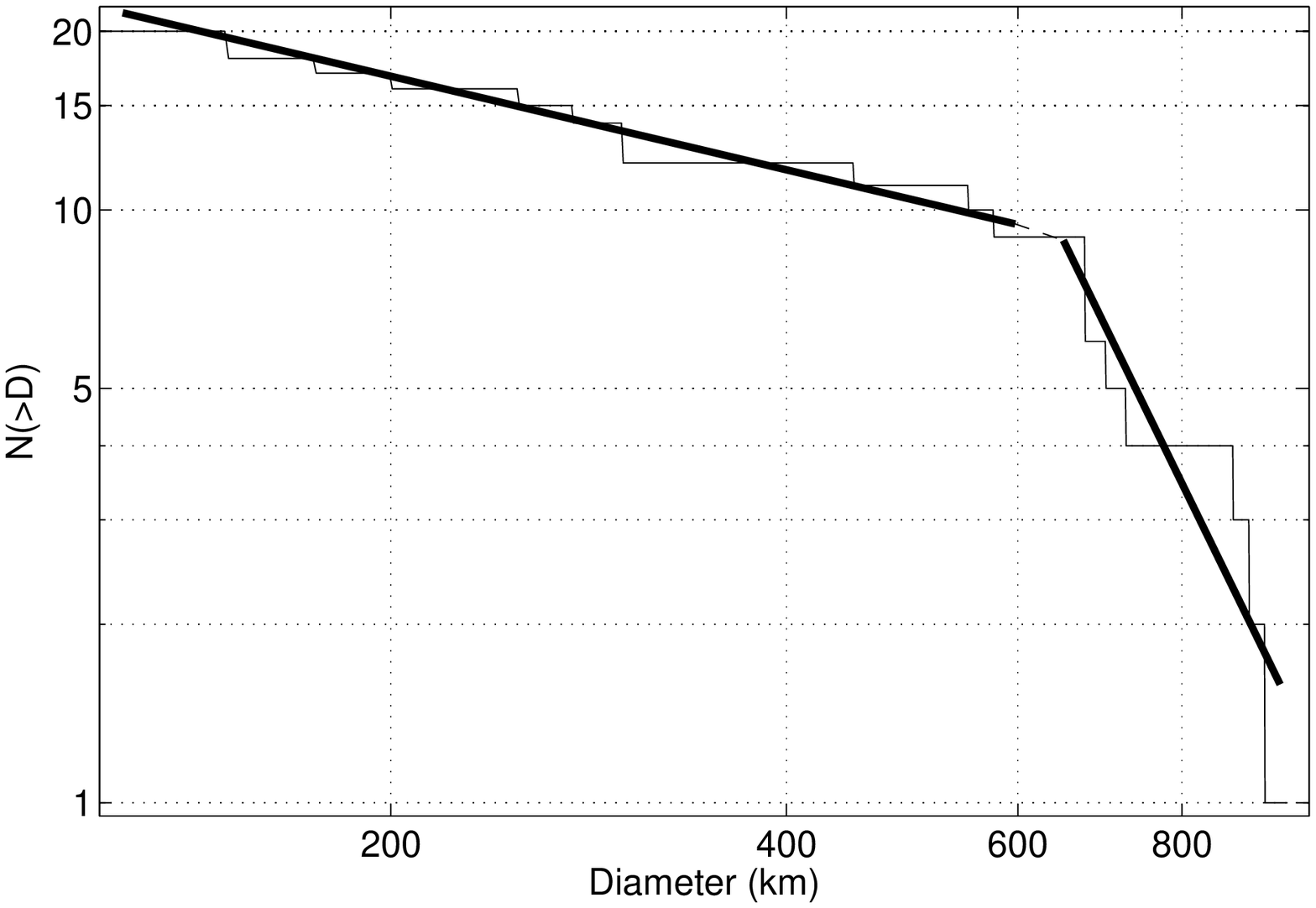}
   \caption{Cumulative size distribution of dynamically hot classicals from
            this work and literature (Table~\ref{collection}).
    The power law has a change between 500 and 700 km. The intermediate
    size classicals have a slope parameter of $q=1.4$.}
   \label{diameter_cumulative}
\end{figure}

The diameter estimates in our sample are ranging from 100 km of 120181 (2003 UR$_{292}$) up to 930 km of 2002 MS$_4$,
which is
larger than previously estimated for it. 
2002 MS$_4$ and 120347 Salacia are 
among the ten largest TNOs
with sizes similar to those of 50000 Quaoar and 90482 Orcus.
The size distribution of hot classicals in
our sample is wider than that of the cold classicals, which are limited to
diameters of 
100-$350$ 
km (Fig.~\ref{diameter_histo}). 
The diameters of eight hot classicals from literature data
(Table~\ref{collection}) are within the same size range as the hot classicals in our sample. 
The cumulative size
distribution of this extended set of 20 hot classicals (Fig.~\ref{diameter_cumulative}) shows two regimes of a
power law distribution with a turning point between 500 and 700 km. The slope of the 
cumulative distribution $N(>D) \propto D^{-q}$ is $q \approx 1.4$ 
for the $100<D<600\ \mathrm{km}$ (N=11) objects.
There are not enough targets for a reliable slope determination in the $D>600\ \mathrm{km}$ regime.
The size distribution is an important property in understanding the processes of planet formation.
Several works have derived it from the LF using simplifying assumptions about common albedo and distance.
\cite{Fraser2010} (2010) reported a slope of the differential size distribution of $2.8 \pm 1.0$
for a dynamically hot TNO population
(38 AU $<$ heliocentric distance $<$ 55 AU and $i>5\degr$).
Our $q+1$ based on a small sample of measured diameters of intermediate-size hot classicals
is compatible with this literature value. The high-$q$ tail at $D>650\ \mathrm{km}$
in Fig.~\ref{diameter_cumulative} indicates a change of slope when the population
transitions from a primordial one to a collisionally relaxed population. Based on LF estimates, this
change in slope for the whole TNO population was expected
at $200-300\ \mathrm{km}$ (\cite{Kenyon2008} based on data from \cite{Bernstein2004}) or at
somewhat larger diameters (\cite{Petit2006}). For Plutinos a change to a steeper slope
occurs at 450 km (\cite{Mommert2012} 2012).

\begin{figure}
   \centering
   \includegraphics[width=10cm]{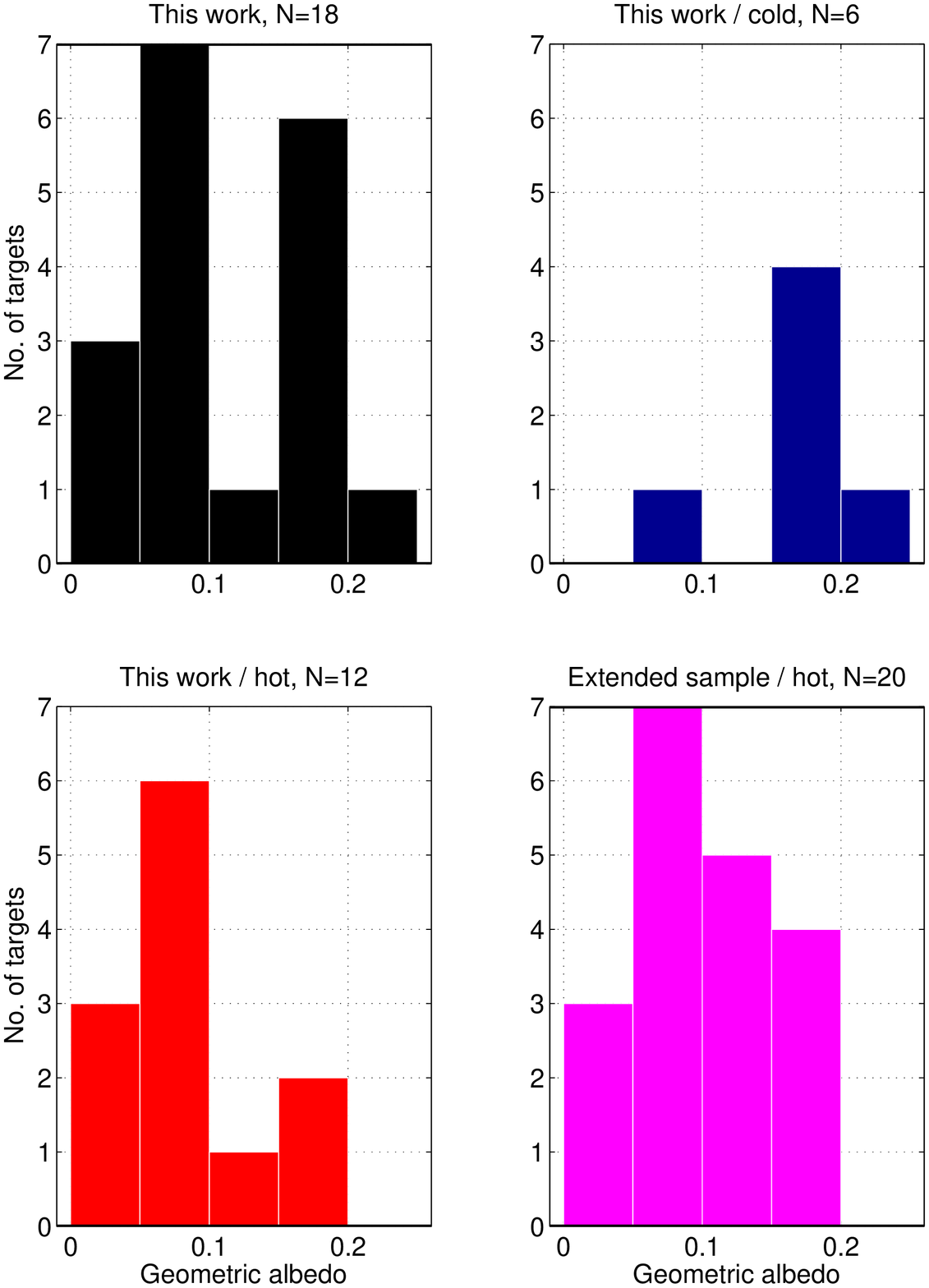}
   \caption{Distribution of geometric albedos from this work (upper left), the cold classicals
of this work (upper right), the hot classicals of this work (lower left), and all hot classicals
including literature results from Table~\ref{collection} (lower right). The bin size is 0.05. The Haumea
family member 55636 (2002 TX$_{300}$) with $p_V=0.88$ is beyond the horizontal scale.}
   \label{albedo_histo}
\end{figure}

The dynamically cold and hot
sub-populations are showing different geometric albedo distributions (Fig.~\ref{albedo_histo})
with the dynamically cold objects
having higher geometric albedos in a narrower distribution.
The average geometric albedo
of the six cold classicals is 
$0.17 \pm 0.04$
(un-weighted 
average and standard deviation).
The highest-albedo object is 88611 Teharonhiawako with
$p_V=0.22$, or possibly 2002 GV$_{31}$ with the lower limit of 0.22.
These findings are compatible with the conclusions of \cite{Brucker2009} (2009) based on \emph{Spitzer} data
that cold classicals have a high albedo, although we do not confirm their extreme geometric albedo of 0.6
for 119951 (2002 KX$_{14}$).

The darkest object in our sample is the dynamically hot target 78799 (2002 XW$_{93}$) with
a geometric albedo of 0.038. The highest-albedo hot classicals are found in the low-$i$ part of
the sub-sample (see Fig.~\ref{D_vs_i}): 138537 (2000 OK$_{67}$) at $i=4.9\degr$ has
$p_{\mathrm{V}}=0.20$ and the inner belt target 120181 (2003 UR$_{\mathrm{292}}$) has a geometric
albedo of 0.16. 
The 12 hot classicals in our sample have, on the average, lower albedos than the cold ones:
$p_{\mathrm{V}}=0.09 \pm 0.05$.
The average
of the combined hot 
classical sub-population of this work and
literature is $0.11 \pm 0.04$ if 55636 (2002 TX$_{300}$) is excluded. 

From the floating-$\eta$ solutions of eight targets with data from both instruments included in the fitted
solution (see Table~\ref{table_results} and Section~\ref{resultsection}) we have the average
$\eta=1.47 \pm 0.43$ (un-weighted). 
Most of these eight targets have $\eta>1$ implying a 
noticeable amount of surface thermal inertia. It should be noted, though, that inferences about surface roughness
and thermal conductivity would require more accurate knowledge of the spin axis orientation and spin period
of these targets.
Our average $\eta$ is consistent with our default value of
$1.20 \pm 0.35$ for fixed $\eta$ fits.
For comparison with other dynamical classes, the average beaming parameter of seven Plutinos is
$\eta=1.11_{-0.19}^{+0.18}$ (\cite{Mommert2012} 2012) and of seven scattered and detached objects $\eta = 1.14 \pm 0.15$
(\cite{SantosSanz2012} 2012). The difference of using a fixed-$\eta=1.47$ instead of fixed-$\eta=1.20$ is that diameters
would increase, on the average, by 10\% and geometric albedos decrease by 16\%. These changes are
within the average relative uncertainties (19\% in diameter and 57\% in geometric albedo) of the fixed-$\eta$ solutions.

\subsection{Correlations}
\label{correlations}
We ran a Spearman rank correlation test (\cite{Spearman1904}) to look for possible 
correlations between the geometric albedo $p_{\mathrm{V}}$, diameter $D$, orbital elements 
(inclination $i$, eccentricity $e$, semimajor axis $a$, perihelion distance 
$q$), beaming parameter $\eta$, visible spectral slope, as well as B-V,
V-R and V-I colors\footnote{From the Minor Bodies in the Outer Solar System database,
\url{http://www.eso.org/~ohainaut/MBOSS}, accesses Nov 2011.}.
The Spearman correlation is a distribution-free test less sensitive to outliers than some other 
more common methods (e.g. Pearson correlation). 
We use a modified form of the test, which takes into account
asymmetric error bars and corrects the significance 
for small numbers statistics. 
The details of our method are
described in \cite{Peixinho2004} and \cite{SantosSanz2012} (2012).
The significance $P$ of a correlation is the probability of
getting a higher or equal correlation coefficient value 
$-1 \le\rho\le 1$ if no correlation existed on the parent population, from which we extracted the sample. 
Therefore, the smaller the P the more unlikely would be to observe a $\rho\neq 0$ if it was indeed equal
to zero, i.e. the greater the confidence
on the presence of a correlation is.
The 99.7\% confidence interval ($3\,\sigma$), or better, corresponds to $P=0.003$, or smaller.
We consider a 'strong correlation' to have $|\rho| \geq 0.6 $,
and a 'moderate correlation' to have $0.3 \leq |\rho|<0.6$.
Selected results from our correlation analysis are presented in Table~\ref{corr_table}
and discussed in the following subsections.

\begin{table*}
\centering
\caption{Selected correlation results (see text).}
\begin{tabular}{llrrlc}
\hline
Variables             & sub-sample                      & Number of    & Correlation            & Significance & Confidence       \\
                      &                                 & data points  & coefficient            &              & limit ($\sigma$) \\
\hline
$D$, $i$              & this work                       & 18           & $0.69_{-0.23}^{+0.14}$  & 0.002       & 3.2 \\
                      & this work / cold                & 6            & $-0.10_{-0.54}^{+0.61}$ & 0.8         & 0.2  \\
                      & this work / hot                 & 12           & $0.82_{-0.22}^{+0.10}$  & 0.0011      & 3.3  \\
                      & this work and previous works       & 26           & $0.65_{-0.15}^{+0.11}$  & 0.0004      & 3.6  \\
                      & this work / hot and previous works & 20           & $0.60_{-0.24}^{+0.16}$  & 0.005       & 2.8  \\
\hline
$D$, $p_{\mathrm{V}}$    & this work                       & 18           & $-0.76^{+0.11}_{-0.08}$ & $0.0002$    & 3.7  \\
                      & this work / cold                & 6            & $-0.40_{-0.41}^{+0.68}$ & 0.4         & 0.8  \\
                      & this work / hot                 & 12           & $-0.62^{+0.28}_{-0.18}$ & $0.03$      & 2.2 \\
                      & this work and previous works       & 26           & $-0.59^{+0.16}_{-0.12}$ & $0.002$     & 3.2 \\
                      & this work / hot and previous works & 20           & $-0.46^{+0.24}_{-0.19}$ & $0.04$      & 2.1 \\
\hline
$H_{\mathrm{V}}$, $i$    & this work                       & 18           & $-0.56_{-0.19}^{+0.27}$ & 0.015       & 2.4 \\
                      & this work / cold                & 6            & $-0.02_{-0.58}^{+0.59}$ & 1.0         & 0.03 \\
                      & this work / hot                 & 12           & $-0.74_{-0.18}^{+0.45}$ & $0.006$     & 2.7 \\ 
\hline
spectral slope, $i$   & this work / hot and previous works & 15           & $-0.65^{+0.32}_{-0.19}$ & 0.011       & 2.6 \\
\hline
\end{tabular}
\label{corr_table}
\end{table*}

\subsubsection{Correlations with diameter}
\label{corr_D}
\begin{figure}
   \centering
   \includegraphics[width=15cm]{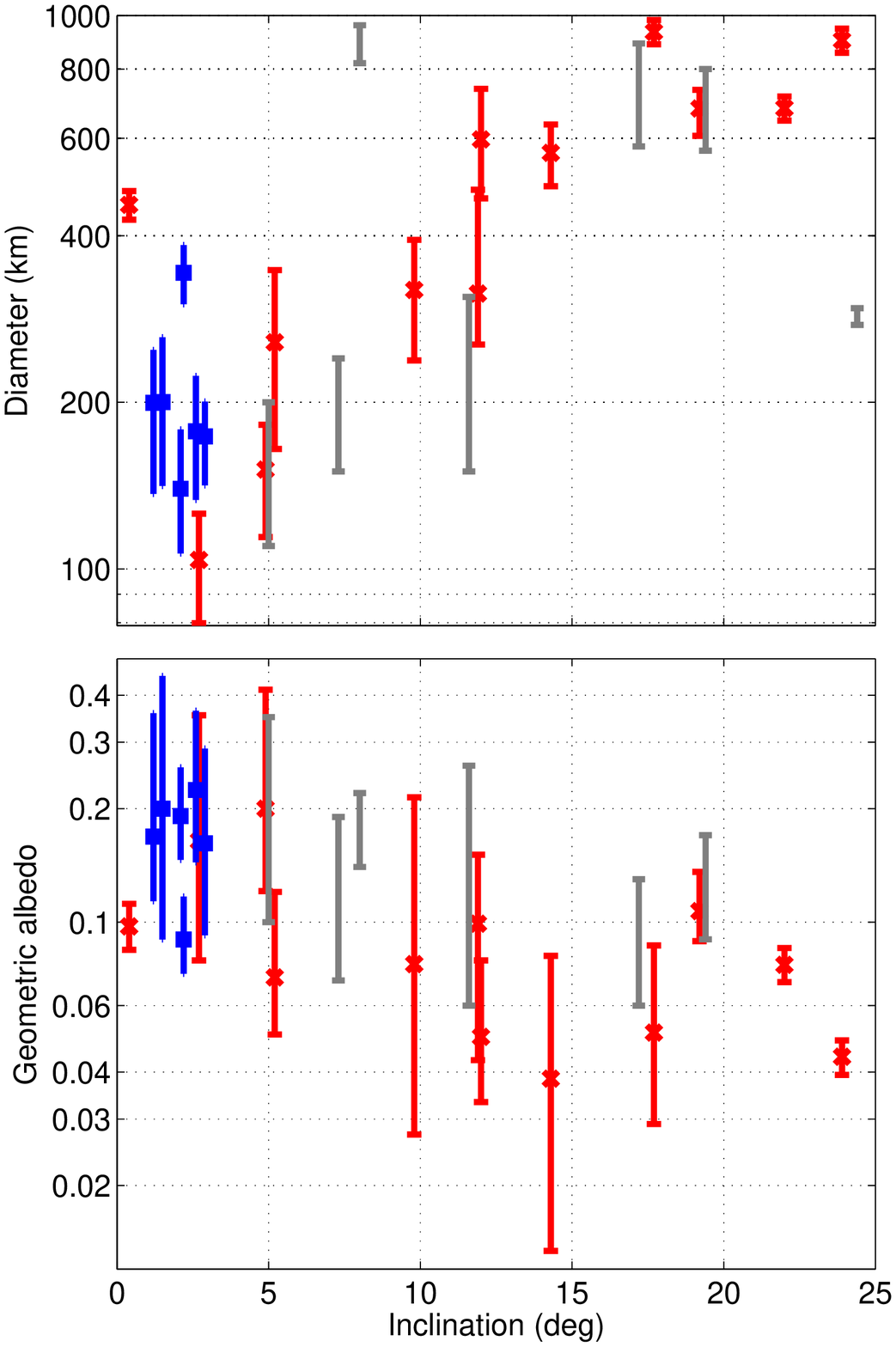}
   \caption{Radiometric diameter as well as geometric albedo vs inclination. The cold classicals of our sample
   are marked with blue squares, hot classicals with red crosses and 
   hot classicals from literature with gray color. The high-albedo target 55636 (2002 TX$_{300}$) is beyond
   the scale.}
   \label{D_vs_i}
\end{figure}

We detect a strong size-inclination correlation in our target sample
(see Fig.~\ref{D_vs_i} and Table~\ref{corr_table}).
When literature targets, all of whom are dynamically
hot, are included in the analysis we get a correlation of similar strength.
Previously this presumable trend has been extrapolated from the correlation between intrinsic brightness 
and inclination (\cite[2001]{Levison2001}).
We see this strong size-inclination correlation also among the hot classicals sub-sample,
but not 
among our cold classicals where we are limited by the small sample size.

Other orbital parameters do not correlate with size. We find no correlation between
size and colors, or spectral slopes, nor
between size and the beaming parameter $\eta$. The possible correlation between size
and geometric albedo is discussed in Section~\ref{corr_pV}.

\subsubsection{Correlations with geometric albedo}
\label{corr_pV}

\begin{figure}
   \centering
   \includegraphics[width=13cm]{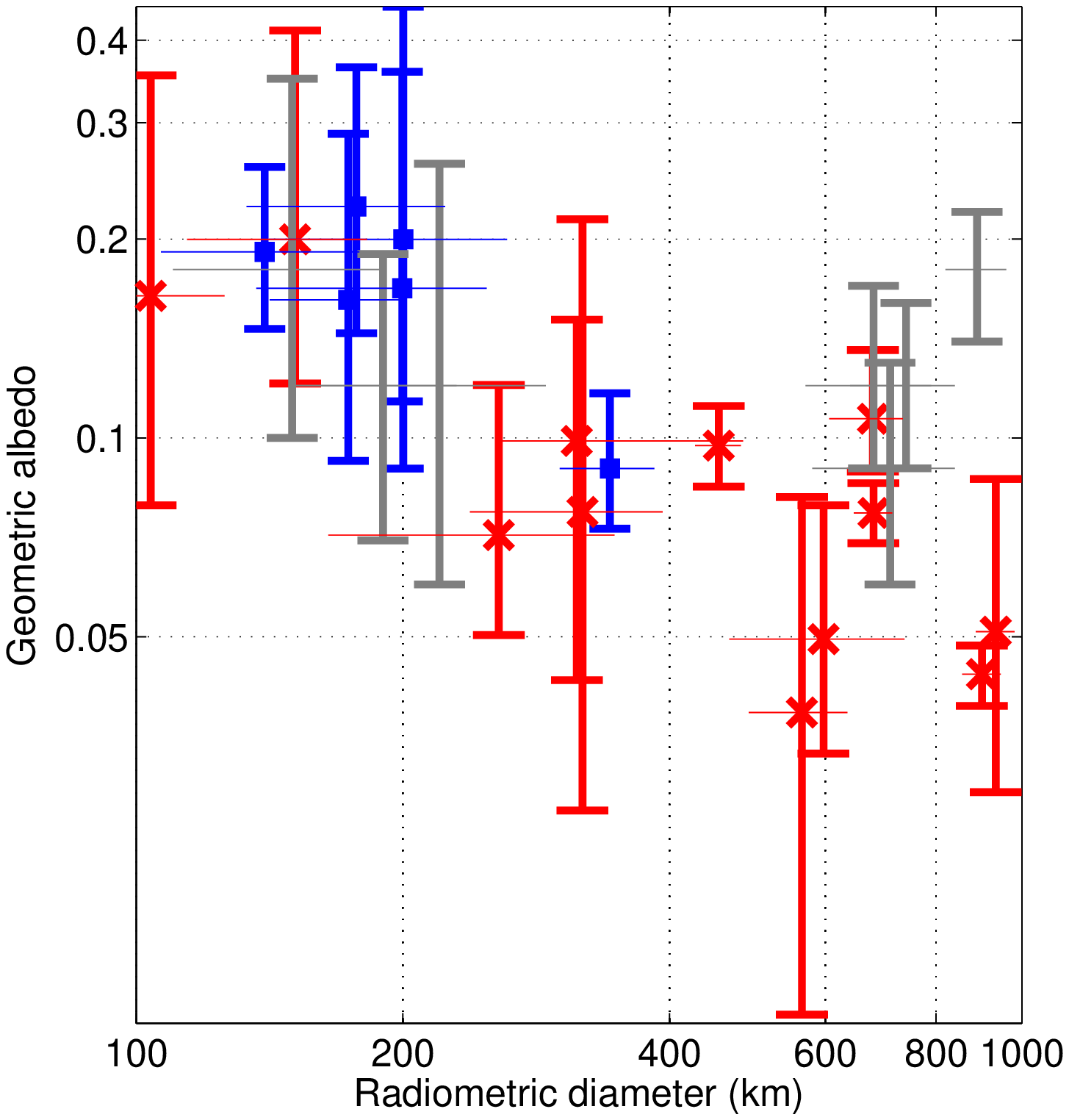}
   \caption{Geometric albedo vs radiometric diameter ({\it blue triangles} 
            = cold classicals, {\it red crosses} = hot classicals from our sample,
   {\it gray} = other hot classicals from literature, see Table~\ref{collection}). }
   \label{corr_fig}
\end{figure}

We find evidence for an anti-correlation between diameter and geometric albedo, both in our sample 
and when combined with other published data of classical TNOs (see Fig.~\ref{corr_fig} and Table~\ref{corr_table}). 
Other dynamical populations with accurately measured diameters/albedos show a different behavior:
there is no such correlation seen among the Plutinos (\cite{Mommert2012} 2012) and a combined sample of 15
scattered-disc and detached objects show a positive correlation between diameter and geometric albedo
at $2.9\,\sigma$ level (\cite{SantosSanz2012} 2012).

As it might be suggested visually by the distribution of diameters of classical TNOs (see Figs.~\ref{diameter_histo} and
\ref{corr_fig}), we have analyzed the possibility of having
two groups with different size-albedo behaviors, separating in size at $D\approx 500$ km regardless of their
dynamical cold/hot membership. We have found no statistical evidence for it.

With our method of accounting for error bars, which tend to `degrade' the correlation values,
geometric albedo does not correlate with
$H_{\mathrm{V}}$, orbital parameters, spectral slopes, colors, nor
beaming parameters $\eta$. Also when the literature targets are added we find
no evidence of correlations.

\subsubsection{Other correlations}
\label{corr_other}

The known correlations between surface color/spectral slope 
and orbital inclination (\cite{Trujillo2002}, \cite[2002]{Hainaut2002}), 
and between intrinsic brightness and inclination (\cite{Levison2001} 2001), 
usually interpreted as a size-inclination correlation, might lead us to conclude there was a consequent 
color/slope-size correlation. Our analysis with measured diameters does not show a correlation neither
with spectral slope nor visible colors, 
as one might expect. Note, however, that we do not possess 
information on the surface colors/slopes of $\sim1/2$ of our targets  
($\sim1/3$ when complemented with other published data) leading to the non-detection of the color/slope-inclination trend, 
which is known to exist among classicals. Thus, a more complete set 
of color/slope data would be required for our targets.
Only when combining the hot sub-sample from this work and literature we see a
non-significant 
anti-correlation 
between slope and inclination (see Table~\ref{corr_table}). We do not find any correlations of
the B-V, V-R and V-I colors with other parameters.

The apparent $H_{\mathrm{V}}$ vs $i$ anti-correlation in our target sample mentioned in Section~\ref{Targetsample}
is almost significant ($2.7\,\sigma$) for our hot sub-population (see Table~\ref{corr_table}).

\subsection{Binaries}
\label{binaries}
Binary systems are of particular scientific interest because
they provide unique constraints on the elusive bulk composition,
whereas all other observational constraints of the composition 
only pertain to the surface of the object.
The sizes of binaries can be constrained
based on the relative brightness difference
of the primary and the secondary components, but only if suitable assumptions about
the relative geometric albedo are made. Alternatively, geometric albedos can be
constrained under certain assumptions about the relative sizes.
The ranges given in the literature are usually
based on the following assumptions: i) the primary and secondary objects are spherical,
ii) the primary and secondary have equal albedos,
and iii) objects have densities within a limited assumed range. 

Six of our targets are binaries with known total mass $m$ 
and brightness difference between the two components $\Delta V$
(see Table~\ref{bin}). Assuming the two components to have 
identical albedos, the latter can be converted into an area
ratio and, assuming spherical shape, a diameter ratio 
$k=D_2/D_1$ (with component diameters $D_1$ and $D_2$). 
Component diameters follow from the measured NEATM diameter 
$D$ (see Table~\ref{table_results}) and $D_2/D_1$:
$D^2 = D_1^2 + D_2^2$ (since $D$ is the area-equivalent system diameter). 
This leads to a ``volumetric diameter''
$D' = \frac{\left( 1+k^3 \right)^{1/3}}{\sqrt{1+k^2}} D$.
The mass densities $\frac{6 m}{\pi D'^3}$ are given in Table~\ref{bin}.
Within the uncertainties, the measured bulk densities scatter around 
roughly 1 g cm$^{-3}$, consistent with a bulk
composition dominated by water ice, as expected for objects in the 
outer Solar System. Significant mass contributions
from heavier materials, such as silicates, are not excluded however, 
and would have to be compensated by significant
amounts of macroporosity. The largest object, 120347 Salacia, 
has a 
bulk density $>1$ g cm$^{-3}$. This could indicate
a lower amount of macroporosity for this object, which is subject to significantly larger gravitational
self-compaction than our other binary targets.

\begin{table}
\centering
\caption{New density estimates for binaries.} 
\begin{tabular}{llccc}
\hline
Target &  Adopted $\Delta$V\tablefootmark{a}  & Mass\tablefootmark{a}                   & Bulk density            \\
       &  (mag)              & ($\times 10^{18}$ kg)  & (g cm$^{-3}$)     \\
\hline
(2001 XR$_{254}$)        & 0.43                     & $4.055 \pm 0.065$ & $1.4_{-1.0}^{+1.3}$       \\
275809 (2001 QY$_{297}$) & 0.20                     & $4.105 \pm 0.038$ & $1.4_{-1.3}^{+1.2}$       \\
79360 Sila              & 0.12\tablefootmark{b} & $10.84 \pm 0.22$  & $0.73 \pm 0.28$ \\
88611 Teharonhiawako    & 0.70                     & $2.445 \pm 0.032$ & $1.14_{-0.91}^{+0.87}$               \\
148780 Altjira          & 0.23                     & $3.986 \pm 0.067$ & $0.63_{-0.63}^{+0.68}$       \\
120347 Salacia          & 2.32                     & $466 \pm 22$      & $1.38 \pm 0.27$   \\
\hline
\end{tabular}
\label{bin}
\tablefoot{{\bf References.}
\tablefoottext{a}{\cite{Grundy2011}.} 
\tablefoottext{b}{\cite{Grundy2012}.}
}
\end{table}

\section{Conclusions}
\label{conclude}
The number of classical TNOs with both the size and the geometric albedo measured radiometrically
is increased by eight from 22 to 30.
Four other targets, which previously had estimated size ranges from the analysis of binary systems,
now have more accurate size estimates.
The number of targets observed
and analysed within the ``TNOs are Cool'' program 
(\cite{Muller2010} 2010, \cite{Lellouch2010} 2010, \cite{Lim2010} 2010,
\cite{SantosSanz2012} 2012, \cite{Mommert2012} 2012) is increased by 18 and
the observation of 79360 Sila disturbed by a background source in 
\cite{Muller2010} (2010) has been re-observed and
analyzed.
Furthermore, three targets which earlier had upper and lower limits 
only based on \emph{Spitzer} data alone
(148780 Altjira, 138537 (2000 OK$_{67}$) and 2001 RZ$_{143}$)
now have accurately estimated diameters and albedos. The new Altjira 
solution is outside of the previous limits
based on \emph{Spitzer} data alone. 
The three PACS data points near the thermal peak are providing 
reliable diameter/albedo solutions,
but in some cases adding \emph{Spitzer} data, especially the 
$24\ \mathrm{\mu m}$ data point in the lower-wavelength regime,
constrains the solution 
and allows smaller error bars and more reliable estimates of the beaming parameter.
Compared to previous works the size estimates of
119951 (2002 KX$_{14}$), and 2002 MS$_4$
have increased. 2002 MS$_4$ (934 km) is similar in size to 50000 Quaoar 
and the refined size of 120347 Salacia 
(901 km) is similar to that of 90482 Orcus.
We find a diameter for 2001 KA$_{77}$, which is approximately half of the previous estimate
(\cite{Brucker2009} 2009), and a geometric albedo approximately 4 times higher. The largest change in estimated
geometric albedo is
with 119951 (2002 KX$_{14}$) from 0.60 to 0.097.

The main conclusions based on accurately measured classical TNOs are:
   \begin{enumerate}
    \item There is a large diversity of objects' diameters and geometric albedos among classical TNOs. 
    \item The dynamically cold targets have higher (average $0.17 \pm 0.04$) and differently distributed albedos
          than the dynamically hot targets ($0.09 \pm 0.05$) in our sample. When extended by seven hot classicals
	  from literature the average is $0.11 \pm 0.04$.
\item Diameters of classical TNOs strongly correlate with orbital inclination in the sample of targets, whose 
      size and geometric albedo have been accurately measured, i.e. low inclination objects are smaller.
      We find no clear evidence of an albedo-inclination trend. 
\item Our data suggests that geometric albedos of classical TNOs 
      anti-correlate with diameter, i.e. smaller objects possess higher albedos.
     \item Our data does not show evidence for correlations between surface colors, 
           or spectral slope, of classical TNOs and their diameters nor with their albedos. 
    \item We are limited by the small sample size of radiometrically
          measured accurate diameters/albedos of dynamically cold classicals (N=6)
          finding no statistical evidences for any correlations.
    \item The cumulative size distribution of hot classicals based on the sample of measured sizes in the range
          of diameters between 100 and 600 km (N=11) has a slope of $q \approx 1.4$.
    \item We determine the bulk densities of six classicals. They scatter around $\sim1$ g cm$^{-3}$.
          The high-mass object 120347 Salacia has a density of $\left( 1.38 \pm 0.27 \right)$ g cm$^{-3}$.
   \end{enumerate}

\begin{acknowledgements}
      We thank Chemeda Ejeta
      for his work in the dynamical classification of the
      targets of the ``TNOs are Cool'' program.
      We acknowledge the helpful efforts of David Trilling in the early planning of this program.

      Part of this work was supported by the German
      \emph{DLR} project numbers 50 OR 1108, 50 OR 0903, 50 OR 0904 and 50OFO 0903. M.~Mommert acknowledges support
      trough the DFG Special Priority Program 1385.
      C. Kiss and A. Pal acknowledge the support of the Bolyai Research Fellowship of the Hungarian Academy of
      Sciences. J.~Stansberry acknowledges support by NASA through an award issued by JPL/Caltech. R.~Duffard
      acknowledges financial support from the MICINN (contract Ram\'on y Cajal). P. Santos-Sanz would like to
      acknowledge financial support by the Centre National de la Recherche Scientifique (CNRS). JLO acknowledges
      support from spanish grants AYA2008-06202-C03-01, P07-FQM-02998 and European FEDER funds.

\end{acknowledgements}


\begin{thebibliography}{}
\bibitem[Altenhoff et al. (2004)]{Altenhoff2004}Altenhoff, W.~J., Bertoldi, F., Menten, K.~M., 2004, A\&A 415, 771.
\bibitem[Barucci et al. (2000)]{Barucci2000}Barucci, M.~A., Romon, J., 
Doressoundiram, A., Tholen, D.~J., 2000, AJ 120, 496.
\bibitem[Barucci et al.]{Barucci2011}Barucci, M.~A., Alvarez-Candal,~A., Merlin, F., et al, 2011, Icarus 214, 297.
\bibitem[Batygin et al. 2011]{Batygin2011}Batygin, K., Brown, M.E. and Fraser, W.C., 2011, AJ 738, 13.
\bibitem[Belskaya et al.]{Belskaya2008}Belskaya, I.N., Levasseur-Regourd, A.-C., 
Shkuratov, Y.G., Muinonen, K., {\it in The Solar
System Beyond Neptune, eds. Barucci, M.A., Boehnhardt, H., Cruikshank, D.P., Morbidelli, A., ISBN 978-0-8165-2755-7}, 115.
\bibitem[Benecchi et al. (2009)]{Benecchi2009}Benecchi, S.D., Noll, K.S., Grundy, W.M., et al., 2009, Icarus 200, 292.
\bibitem[Benecchi et al.]{Benecchi2011}Benecchi, S.D., Noll, K.S., Stephens, D.C. et al., 2011, Icarus 213, 693.
\bibitem[Bernstein et al. 2004]{Bernstein2004}Bernstein, G.M., Trilling, D.E., Allen, R.L., 2004, AJ 128, 1364.
\bibitem[Bessell et al. 1998]{Bessell1998}Bessell, M.S., Castelli, F., Plez, B., 1998, A\&A 333, 231.
\bibitem[Boehnhardt et al. (2001)]{Bohnhardt2001}Boehnhardt, H., Tozzi, G.P., Birkle, K., 2001, A\&A 378, 653.
\bibitem[Bowell et al. 1989]{Bowell1989}Bowell, E., Hapke, B., Domingue, D., et al., 
1989, {\it in Asteroids II}, University of Arizona Press.
\bibitem[Brucker et al.]{Brucker2009}Brucker, M.J., Grundy, W.M., Stansberry, J.A., et al., 2009, Icarus 201, 284.
\bibitem[Chiang et al. 1999]{Chiang1999}Chiang, E.I, Brown, M.E., 1999, AJ 118, 1411.
\bibitem[Davies et al. (2000)]{Davies2000}Davies, J.K., Green, S., McBride, N., et al., 2000, Icarus 146, 253.
\bibitem[Delsanti et al. (2001)]{Delsanti2001}Delsanti, A.C., B{\"o}hnhardt, H., Barrera, L., et al., A\&A 380, 347.
\bibitem[DeMeo et al. (2009)]{DeMeo2009}DeMeo, F.E., Fornasier, S., Barucci, M.A., et al., 2009, A\&A 493, 283.
\bibitem[Duffard et al. 2009]{Duffard2009}Duffard, R., Ortiz, J.L., Thirouin, A. et al., 2009, A\&A 505, 1283.
\bibitem[Doressoundiram et al. (2002)]{Doressoundiram2002}Doressoundiram, A., 
Peixinho, N., de Berg, C., et al., 2002, AJ 124, 2279.
\bibitem[Doressoundiram et al. (2005)]{Doressoundiram2005}Doressoundiram, A., 
Peixinho, N., Doucet, C., et al., 2005, Icarus 174, 90.
\bibitem[Doressoundiram et al. (2007)]{Doressoundiram2007}Doressoundiram, A., 
Peixinho, N., Moullet, A., et al., 2007, AJ, 134, 2186.
\bibitem[Dotto et al. (2008)]{Dotto2008}Dotto, E., Perna, D., Barucci, M.A., et al., 2008, A\&A 490, 829.
\bibitem[Elliot et al.]{Elliot2005}Elliot, J.L., Kern, S.D., Clancy, K.B., et al., 2005, AJ 129, 1117.
\bibitem[Elliot et al. (2010)]{Elliot2010}Elliot, J.L., Person, M.~J., Zuluaga, C.~A. et al. 2010, Nature 465, 897.
\bibitem[Engelbracht et al. (2007)]{Engelbracht2007}Engelbracht, C.~W., 
Blaylock, M., Su, K.~Y.~L. et al., 2007, \pasp \ 119, 994.
\bibitem[Fornasier et al. (2009)]{Fornasier2009}Fornasier, S., Barucci, M.~A., de Bergh, C. et al., 2009, A\&A 508, 457.
\bibitem[Fraser and Brown]{Fraser2010b}Fraser, W.C. and Brown, M.E., 2010, ApJ 714, 1547.
\bibitem[Fraser et al.]{Fraser2010}Fraser, W.C., Brown, M.E. and Schwamb, M.E., 2010, Icarus 210, 944.
\bibitem[Fulchignoni et al. (2008)]{Fulchignoni2008}Fulchignoni, M., 
Belskaya, I., Barucci, M.~A. et al., 2008 {\it in The Solar
System Beyond Neptune, eds. Barucci, M.A., Boehnhardt, H., Cruikshank, D.P., Morbidelli, A.}, 181.
\bibitem[Giorgini et al. 1996]{Giorgini1996}Giorgini, J.D., Yeomans, D.K., 
Chamberlin, A.B., et al., 1996, Bulletin of AAS 28(3), 1158.
\bibitem[Gladman et al. (2008)]{Gladman2008}Gladman, B., Marsden, B.G., VanLaerhoven, Ch., 2008, {\it in The Solar
System Beyond Neptune, eds. Barucci, M.A., Boehnhardt, H., Cruikshank, D.P., Morbidelli, A.}, 43.
\bibitem[Gordon et al. (2007)]{Gordon2007}Gordon, K.~D., Engelbracht, C.~W., Fadda, D., 2007, \pasp \ 119, 1019.
\bibitem[Grundy et al. 2005]{Grundy2005}Grundy, W.M., Noll, K.S., Stephens, D.C., 2005, Icarus, 176, 184.
\bibitem[Grundy et al. (2009)]{Grundy2009}Grundy, W.M., Noll, K.S., Buie, M.W. et al., 2009, Icarus 200, 627.
\bibitem[Grundy et al. (2011)]{Grundy2011}Grundy, W.M., Noll, K.S., Nimmo, F. et al., 2011, Icarus 213, 678.
\bibitem[Grundy et al. (2012)]{Grundy2012}Grundy, W.M., Benecchi, S.~D., 
Rabinowitz, D.~L. et al., 2012, submitted to Icarus.
\bibitem[Hainaut and Delsanti]{Hainaut2002}Hainaut, O., Delsanti, A., 2002, A\&A 389, 641, updated database
\url{http://www.eso.org/~ohainaut/MBOSS/}, accessed July 2011.
\bibitem[Harris 1998]{Harris1998}Harris, A. W., 1998, Icarus 131, 291.
\bibitem[Harris 2006]{Harris2006}Harris, A. W., 2006, 
{\it in Asteroids, Comets, Meteors, Proceedings IAU Symposium No. 229, 2005, eds. 
D. Lazzaro, S. Ferraz-Mello and J.~A. Fern{\'a}ndez}.
\bibitem[Hayes 1985]{Hayes1985}Hayes, D.~S., 1985, {\it in IAU symposium 111, eds. Hayes, D.S. et al.}, 225.
\bibitem[Hovis and Callahan 1966]{Hovis1966}Hovis, W.~A., Callahan, W.~R., 1966, J. Opt. Soc. Amer., 56, 639.
\bibitem[Jewitt and Luu 1993]{Jewitt1993}Jewitt, D., Luu, J., 1993, Nature 362, 730.
\bibitem[Kenyon et al. 2008]{Kenyon2008}Kenyon,~S.~J., Bromley,~B.~C., 
O'Brien,~D.~P., Davis,~D.~R., 2008, {\it in The Solar
System Beyond Neptune, eds. Barucci, M.A., Boehnhardt, H., Cruikshank, D.P., Morbidelli, A.}, 293.
\bibitem[Kiss et al. 2005]{Kiss2005}Kiss, Cs., Klaas, U., Lemke, D., 2005, A\&A 430, 343.
\bibitem[Lagerros 1996]{Lagerros1996}Lagerros, J.S.V., 1996, A\&A 310, 1011.
\bibitem[Lebofsky et al. 1986]{Lebofsky1986}Lebofsky, L.A., Sykes, M.V. Tedesco, E.F. et al., 1986, Icarus 68, 239.
\bibitem[Lebofsky and Spencer, 1989]{Lebofsky1989}Lebofsky, L.A., Spencer, J.R., 1989, 
{\it in Asteroids II, eds. Binzel, R.P., Gehrels, T., Matthews, M.S.}, Arizona University Press, 128.
\bibitem[Lellouch et al.]{Lellouch2010}Lellouch, E., Kiss, Cs., Santos-Sanz, P. et al., 2010, A\&A 518, L147.
\bibitem[Levison and Stern]{Levison2001}Levison, H.F., Stern, S.A., 2001, AJ 121, 1730.
\bibitem[Levison et al. 2008]{Levison2008}Levison, H.F., Morbidelli A., VanLaerhoven, Ch. et al., 2008, Icarus 196, 258.
\bibitem[Lim et al.]{Lim2010}Lim, T.L., Stansberry, J., M\"uller, Th. et al, 2010, A\&A 518, L148.
\bibitem[Matson (1971)]{Matson1971}Matson, D.L., 1971, PhD thesis, California Institute of Technology.
\bibitem[Mommert et al.]{Mommert2012}Mommert, M., Harris, A.~W., Kiss, C., et al., 2012, accepted for publication in A\&A.
\bibitem[Morbidelli et al. 2008]{Morbidelli2008}Morbidelli, A., Levison, H.F., Gomes, R., 2008, {\it in The Solar
System Beyond Neptune, eds. Barucci, M.A., Boehnhardt, H., Cruikshank, D.P., Morbidelli, A.}, 275.
\bibitem[Moro-Mart\'in et al. 2008]{MoroMartin2008}Moro-Mart\'in, A., Wyatt, M.C., 
Malhotra, R., Trilling, D.E., 2008, {\it in The Solar
System Beyond Neptune, eds. Barucci, M.A., Boehnhardt, H., Cruikshank, D.P., Morbidelli, A.}, 465.
\bibitem[Mueller et al. (2011)]{Mueller2011}Mueller, M., Delbo, M., Hora, J.L. et al., 2011, AJ 141, 109.
\bibitem[M\"uller and Lagerros 1998]{Muller1998}M\"uller, T.~G., Lagerros, J.~S.~V., 1998, A\&A 338, 340.
\bibitem[M\"uller et al. 2009]{Muller2009}M\"uller, T.~G., Lellouch, E., 
B{\"o}hnhardt, H. et al., 2009, Earth Moon and Planets, 105, 209.
\bibitem[M\"uller et al.]{Muller2010}M\"uller, Th., Lellouch, E., Stansberry, J., et al., 2010, A\&A 518, L146.
\bibitem[Noll et al. 2008]{Noll2008}Noll, K.S., Grundy, W.M., Stephens, D.C., et al., 2008, Icarus 194, 758.
\bibitem[Osip et al. (2003)]{Osip2003}Osip, D.J., Kern, S.D., Elliot, J.L., 2003, Earth, Moon, Planets 92, 409.
\bibitem[PACS AOT release note 2010]{PACSrelnote}PACS AOT Release Note: PACS Photometer Point/Compact Source Mode, 2010,
PICC-ME-TN-036, Version 2.0, custodian Th. M\"uller, 
\url{http://herschel.esac.esa.int/twiki/bin/view/Public/PacsCalibrationWeb}.
\bibitem[PACS photometer -- Point Source Flux Calibration 2011]{PACScal2011}PACS photometer -- 
Point Source Flux Calibration 2011,
PICC-ME-TN-037, Version 1.0, \url{http://herschel.esac.esa.int/twiki/bin/view/Public/PacsCalibrationWeb}.
\bibitem[PACS photometer PSF 2010]{PSF2010}PACS photometer point spread function, 2010, PICC-ME-TN-033, Version 1.01,
custodian D. Lutz, \url{http://herschel.esac.esa.int/twiki/bin/view/Public/PacsCalibrationWeb}.
\bibitem[Peixinho et al. (2004)]{Peixinho2004}Peixinho, N., Boehnhardt, H., Belskaya, I. et al., 2004, Icarus 170, 153.
\bibitem[Peixinho et al. 2008]{Peixinho2008}Peixinho, N., Lacerda, P. and Jewitt, D., 2008, AJ 136, 1837.
\bibitem[Perna et al. (2010)]{Perna2010}Perna, D., Barucci, M.A., Fornasier, S., 2010, A\&A 510, A53.
\bibitem[Petit et al 2006]{Petit2006}Petit, J-M., Holman,~M.~J., Gladman, B., et al., 2006, {\mnras} 365, 429.
\bibitem[Petit et al 2008]{Petit2008}Petit, J-M., Kavelaars, J.J., Gladman, B., Loredo, T., 2008, {\it in The Solar
System Beyond Neptune, eds. Barucci, M.A., Boehnhardt, H., Cruikshank, D.P., Morbidelli, A.}, 71.
\bibitem[Petit et al. 2011]{Petit2011}Petit, J-M., Kavelaars, J.J., Gladman, B., et al., 2011, AJ 142, 142.
\bibitem[Pilbratt et al. 2010]{Pilbratt2010}Pilbratt, G.L., Riedinger, J.R., Passvogel, T., 2010, A\&A 518, L1.
\bibitem[Poglitsch et al. 2010]{Poglitsch2010}Poglitsch, A., Waelkens, C., Geis, N. et al., 2010, A\&A 518, L2.
\bibitem[Rabinowitz et al. (2007)]{Rabinowitz2007}Rabinowitz, D.L., Schaefer, B.E., Tourtellotte, S.W., 2007, AJ 133, 26.
\bibitem[Rieke et al. 2004]{Rieke2004}Rieke, G.H., Young, E.T., Engelbracht, C.W. et al. 2004, ApJS 154, 25.
\bibitem[Romanishin and Tegler]{Romanishin2005}Romanishin, W., Tegler, S.C., 2005, Icarus 179, 523.
\bibitem[Santos-Sanz et al.]{SantosSanz2009}Santos-Sanz, P., Ortiz, J.L., Barrera, L., Boehnhardt, H., 2009, A\&A 494, 693.
\bibitem[Santos-Sanz et al.]{SantosSanz2012}Santos-Sanz, Lellouch, E., Fornasier, S., 
et al., 2012, accepted for publication in A\&A.
\bibitem[Schmitt et al. 1998]{Schmitt1998}Schmitt, B., Quirico, E., Trotta, F., Grundy, W.~M., 1998, 
{\it in Solar System Ices, Based on reviews presented at the 
international symposium "Solar system ices" held in Toulouse, France, 
on March 27-30, 1995, eds. Schmitt, B., de Bergh, C., Festou, M.}, 
Dordrecht Kluwer Academic Publishers, Astrophysics and space science library (ASSL) Series, 227, ISBN 0792349024, 199.
\bibitem[Sheppard (2007)]{Sheppard2007}Sheppard, S.S., 2007, AJ 134, 787.
\bibitem[Sheppard and Jewitt]{Sheppard2002}Sheppard, S.S., Jewitt, D.C., 2002, AJ 124, 1757.
\bibitem[Spearman 1904]{Spearman1904}Spearman, C., 1904, Am. J. Psychol, 57, 72.
\bibitem[Spencer et al. 1989]{Spencer1989}Spencer, J.R., Lebofsky, L.A., Sykes, M.V., 1989, Icarus 78, 337.
\bibitem[Stansberry et al. 1996]{Stansberry1996}Stansberry, J.~A., Pisano, D.~J., 
Yelle, R.~V., 1996, Planet. Space Sci., 44, 945.
\bibitem[Stansberry et al. (2007)]{Stansberry2007}Stansberry, J., 
Gordon, K.~D., Bhattacharya, B. et al., 2007, {\pasp} 119, 1038.
\bibitem[Stansberry et al.]{Stansberry2008}Stansberry, J., Grundy, W., Brown, M., et al., 2008, {\it in The Solar
System Beyond Neptune, eds. Barucci, M.A., Boehnhardt, H., Cruikshank, D.P., Morbidelli, A.}, 161.
\bibitem[Stansberry et al. (2012)]{Stansberry2012}Stansberry, J.~A., Grundy, W.~G., 
M\"uller, M. et al., 2012, {\it accepted for publication in Icarus}.
\bibitem[Stetson 1987]{Stetson1987}Stetson, P.B., 1987, PASP 99, 191.
\bibitem[Tegler and Romanishin (2000)]{Tegler2000}Tegler, S.~C., Romanishin, W., 2000, Nature 407, 979.
\bibitem[Thirouin et al.]{Thirouin2010}Thirouin, A., Ortiz, J.L., Duffard, R., 2010, A\&A 522, A93.
\bibitem[Thirouin et al.]{Thirouin2012}Thirouin, A., Ortiz, J.L., 
Campo Bagatin, A. et al, 2012, {\it submitted to {\mnras}}.
\bibitem[Trujillo and Brown 2002]{Trujillo2002}Trujillo, C.A. and Brown, M.E., 2002, ApJ 566, L125.
\bibitem[Tsiganis et al. 2005]{Tsiganis2005}Tsiganis, K., Gomes, R., 
Morbidelli, A. and Levison, H.F., 2005, Nature 435, 459.
\bibitem[Veeder et al. 1989]{Veeder1989}Veeder, G.J., Hanner, M.S., Matson, D.L., 1989, AJ 97, 1211.
\bibitem[Werner et al. 2004]{Werner2004}Werner, M.W., Roellig, T.L., Low, F.J. et al., 2004, ApJS 154, 1.
\bibitem[Wyatt 2008]{Wyatt2008}Wyatt, M.C., 2008, ARA\&A 46, 339.
\end{thebibliography}
\end{document}